\documentclass[%
 reprint,
 amsmath,amssymb,
 aps, nofootinbib
]{revtex4-1}

\usepackage{graphicx}
\usepackage{bm}
\usepackage[normalem]{ulem}
\usepackage[dvipsnames, usenames]{xcolor}
\begin{document}

\title{Radiation from Global Topological Strings using Adaptive Mesh Refinement: Methodology and Massless Modes}

\author{Amelia Drew}
 \email{a.drew@damtp.cam.ac.uk}
\author{E.P.S. Shellard}%
 \email{epss@damtp.cam.ac.uk}
\affiliation{%
 Centre for Theoretical Cosmology, Department of Applied Mathematics and Theoretical Physics,
University of Cambridge, Wilberforce Road, Cambridge CB3 0WA, United Kingdom
}%

\date{March 10, 2022}

\begin{abstract}

We implement adaptive mesh refinement (AMR) simulations of global topological strings using the public numerical relativity code, GRChombo. We perform a quantitative investigation of the dynamics of single sinusoidally displaced string configurations, studying a wide range of string energy densities $\mu \propto \ln{\lambda}$, defined by the string width parameter $\lambda$ over two orders of magnitude.  We investigate the resulting massless (Goldstone boson or axion) radiation signals, using quantitative diagnostic tools to determine the eigenmode decomposition.  Given analytic radiation predictions, we compare the oscillating string trajectory with a backreaction model accounting for radiation energy losses, finding excellent agreement. We establish that  backreaction decay is accurately characterised by the inverse square of the amplitude being proportional to the inverse tension $\mu$ for $3\lesssim \lambda \lesssim 100$.  We conclude that analytic radiation modelling in the thin-string (Nambu-Goto) limit provides the appropriate cosmological limit for global strings. We contextualise these results with respect to axions and gravitational waves produced by cosmic string networks.

\end{abstract}

\maketitle

\section{\label{Introduction}Introduction}

The existence of topological strings is a fundamental prediction of many physically motivated field theories \cite{Kibble1976}, from grand-unified (GUT) models to superstring theory, and has a wide variety of cosmological consequences (for a review, see \cite{Vilenkin:2000jqa}). They usually arise as a result of a symmetry-breaking phase transition, which may have occurred in the early Universe as it cooled below a critical temperature.  The simplest model is the breaking of a $U(1)$ symmetry with a single complex scalar field to create so-called `global' strings with a long-range Goldstone boson or axion field.  A key motivation is offered by the Peccei-Quinn $U_{\rm PQ}(1)$ symmetry introduced to solve the strong CP problem of QCD \cite{Peccei1977a}; when $U_{\rm PQ}(1)$ is broken, axion strings are created which are a potential source of dark matter axions \cite{Davis1986}. 
 
The evolution of cosmological strings has been extensively studied using large-scale numerical simulations. However, there is a vast difference in scale between the typical string width $\delta$ and the string curvature scale $\Lambda$ (usually set by the Hubble radius $R \lesssim H^{-1}$, where $H$ is the Hubble scale). This is characterised by the ratios $\ln{R/\delta \sim 70}$ and $\ln{R/\delta \sim 100}$ for QCD axion and GUT scale strings respectively.  This poses a very significant computational challenge. 

To date, two numerical methods have primarily been used for string simulations in an expanding background. The first uses the Nambu-Goto approximation which assumes that the radius  of  curvature  of  the  string  is much  greater  than  its  thickness, so that the string effectively has zero width  \cite{Albrecht:1984xv,Bennett:1989ak,Allen:1990tv,BlancoPillado:2011dq}. This one-dimensional approach achieves a wide dynamic range but does not directly couple to the long-range radiation fields, nor calculate their backreaction effects. The second method numerically evolves the field equations of motion in three dimensions, incorporating the physical effects of radiation \cite{Harari:1987ht,Davis:1989nj,Battye:1994au,Yamaguchi:1998gx}. However, typical field theory simulations can only probe ratios $\ln{R/\delta} \lesssim 8$ (e.g. \cite{Gorghetto:2021}), so struggle to accurately resolve the string cores in a realistic cosmological context. In order to achieve sufficient dynamic range in an expanding universe, the string width in field theory simulation is typically fixed at finite comoving width (see e.g. \cite{Vincent:1997cx,Moore:2001px,Hiramatsu:2012gg,Moore:2017ond,Gorghetto:2018myk,Kawasaki:2018bzv,Hindmarsh:2019csc}). These two very different approximations yield significant quantitative discrepancies regarding the typical string network density and the primary decay mechanism of radiating strings.  This has important consequences, generating a range of predicted cosmic string gravitational wave signatures (see e.g. \cite{Abbott2018}) and a lack of consensus on the mass of the dark matter axion (see e.g. the review \cite{Marsh:2015xka}).

In this paper, we address this issue using the computational technique, adaptive mesh refinement (AMR).  AMR algorithms enable us to simulate cosmic string evolution whilst adapting the size of the numerical grid to the local scale of the problem, using finer resolution at the string core. This opens up the possibility for greater dynamic range, reducing the need to approximate strings as having either zero or fixed comoving width.

In Section \ref{Theory}, we outline the theory of global cosmic/axion strings and their key radiative modes, discussing the discrepancies between current numerical simulations in the literature. In Section \ref{AMR}, we introduce AMR and describe how it can be applied in the context of global strings, along with the other numerical techniques used to set up our simulations. In Section \ref{Massless}, we present our main results for massless radiative modes, investigating their spectral content and making direct comparisons to analytic radiation calculations. Finally, we conclude with the main implications and discuss future work in Section \ref{Conclusion}.

\section{Theory of Global Topological Strings}\label{Theory}

\subsection{Global $U(1)$ Field Theory}

In this section, we outline the theory of global cosmic strings, including their evolution equations and radiation.  We follow closely the outline of \cite{Vilenkin:2000jqa}, where further details can also be found.

Cosmic strings are topological defects that arise due to symmetry breaking within certain field theories. Symmetry breaking may have occurred as a result of a phase transition in the early Universe when the temperature cooled below some critical value, analagous to a ferromagnet cooling past its Curie point \cite{Kibble1976}. `Global' cosmic strings refer to the simplest case, a $U(1)$ symmetry breaking of a complex scalar field $\varphi$. 

We consider the Goldstone model with Lagrangian density given by
\begin{equation}
    \mathcal{L} = (\partial_\mu\bar{\varphi})(\partial^\mu\varphi) - V(\varphi)
    \label{GoldstoneLagrangian}
\end{equation}
and 
\begin{equation}\label{potential}
V(\varphi) = \frac{1}{4}\lambda(\bar{\varphi}\varphi - \eta^2)^2\,.
\end{equation}
The symmetry breaking scale is set by the constant $\lambda$, and the complex scalar field $\varphi$ is given by
\begin{equation}\label{complex}
\varphi = \phi_1 + i\phi_2\,,
\end{equation}
where $\phi_{1,2}$ indicate the real and imaginary parts. The Euler-Lagrange equations are given by 
\begin{equation}
    \frac{\partial^2\phi_{1,2}}{\partial t^2} -
    \nabla^2\phi_{1,2} + \frac{\lambda}{2}\phi_{1,2}(|\varphi|^2 - \eta^2) = 0\,,
     \label{EL}
\end{equation}
where $\eta$ is a constant.

The Goldstone Lagrangian (\ref{GoldstoneLagrangian}) is invariant under the symmetry transformation 
\begin{equation}
\varphi(x) \rightarrow e^{i\alpha}\varphi(x)\,,
\label{vacuum}
\end{equation}
where $\alpha$ is an angle independent of spatial location.  If we apply this transformation to the lowest energy vacuum state
\begin{equation}
    \langle0|\varphi|0\rangle = \eta e^{i\theta}\,
\end{equation}
where $\theta$ is the complex phase $\tan^{-1}(\phi_2/\phi_1)$, a different expectation value
\begin{equation}
    \langle0|\varphi|0\rangle = \eta e^{i(\theta + \alpha)}
    \label{vacuumbrokensym}
\end{equation}
is obtained, spontaneously breaking the symmetry.

\subsection{Structure of Global Strings}\label{globalstrings}

\begin{figure}
    \centering
    \includegraphics[width=\linewidth]{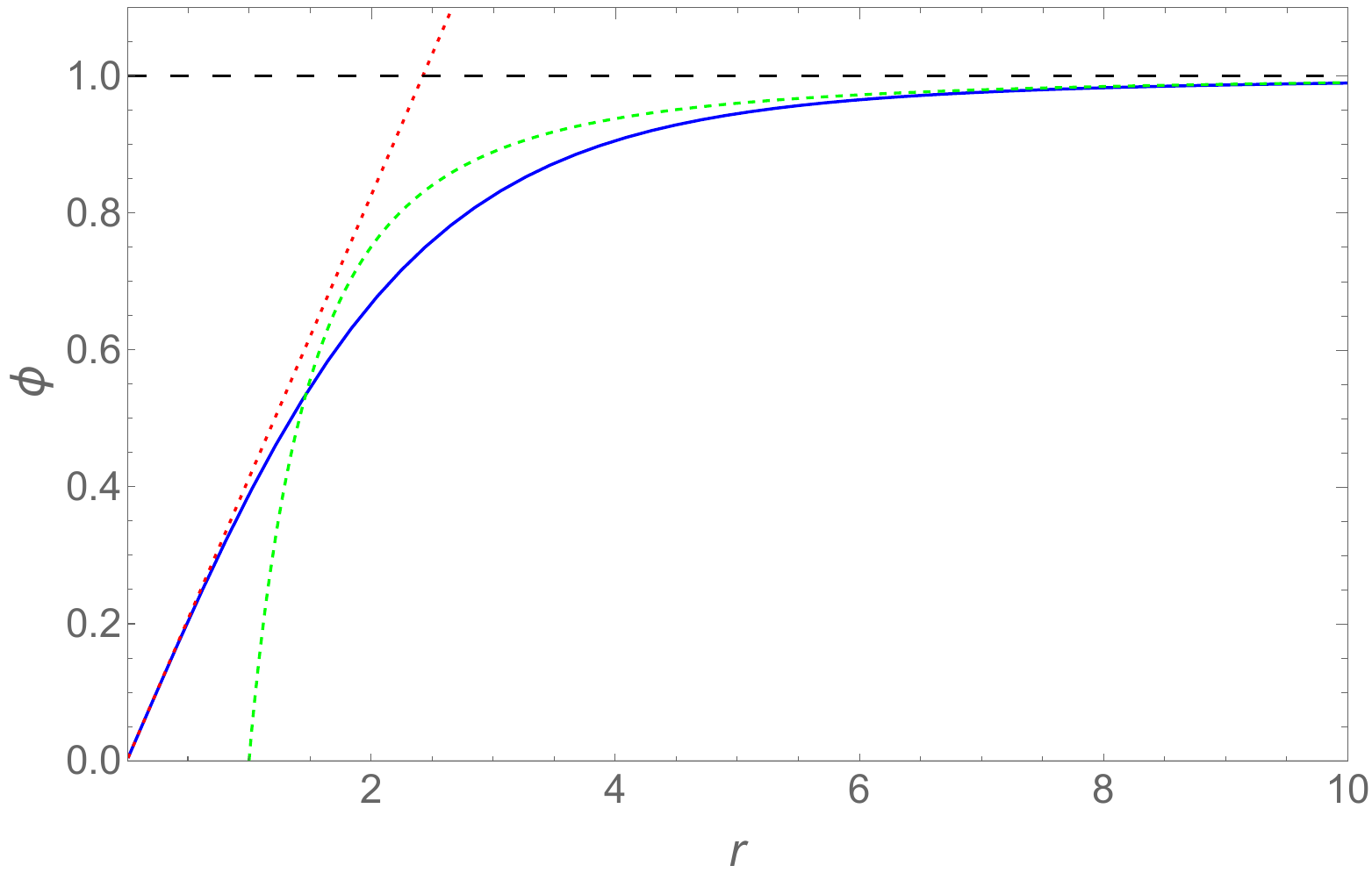}
    \caption{Radial profile of a global cosmic string $\phi(r)$ for $\lambda=1$ with approximate width $\delta \approx 1$ (blue solid). Also plotted are the asymptotic regimes $\phi = 0.412 \,r $ for $r\rightarrow 0$ (red dotted) and $1- r^{-2}$ as $r\rightarrow\infty$ (green dashed).}
    \label{StringCrossSection}
\end{figure}

The possibility of non-trivial windings about the circular vacuum topology leads to the existence of vortex solutions in two dimensions or line-like global strings in three dimensions. 
To find the initial field configuration for strings, we postulate the static \textit{ansatz} solution to (\ref{EL})
\begin{equation}
    \varphi(r, \theta) =  \phi(r)e^{in_w\theta} \label{phi}\,,
\end{equation}
where $\phi = |\varphi|$ and $n_w$ is the topological winding number (an integer). This radial \textit{ansatz} can be substituted into the static part of the  Euler-Lagrange equations (\ref{EL}) to yield an ordinary differential equation for $\phi(r)$
\begin{equation}
    \frac{d^2 \phi}{d r^2} + \frac{1}{r}\frac{d\phi}{dr} - \frac{n_w^2\phi}{r^2} - \frac{\lambda}{2}\phi(\phi^2 - \eta^2) = 0\,,
    \label{radialeqn}
\end{equation}
subject to the boundary conditions $\phi(0)=0$ at the string core and $\phi(r)\rightarrow \eta$ as $r\rightarrow \infty$.     The radial equation (\ref{radialeqn}) can be solved numerically to obtain the string cross-section $\phi(r)$, plotted in Figure \ref{StringCrossSection} for the single winding $n_w=1$ string. This simple solution allows us to infer the width of the string core, defined to be  
\begin{equation}
\delta \approx m_H^{-1} \equiv  (\sqrt\lambda\,\eta)^{-1}\,,
\end{equation}
where $m_H^{-1}$ is the Compton wavelength of the massive particle (see Section \ref{masslessmassive} for further discussion).  As can be seen from Figure \ref{StringCrossSection}, the `half width' of the string core is larger than this, given by $\delta_{1/2}=1.35\,\delta$  where $\phi=0.5$ (or $1.68\,\delta$ using the same criteria with the energy density). In subsequent calculations, we perform a rescaling to set $\eta=1$ (using  $\phi \rightarrow  \phi/\eta$ and $r\rightarrow \eta r$), but retain $\lambda$ as a free parameter to modify the string width.

\begin{figure}
    \centering
    \includegraphics[width=\linewidth]{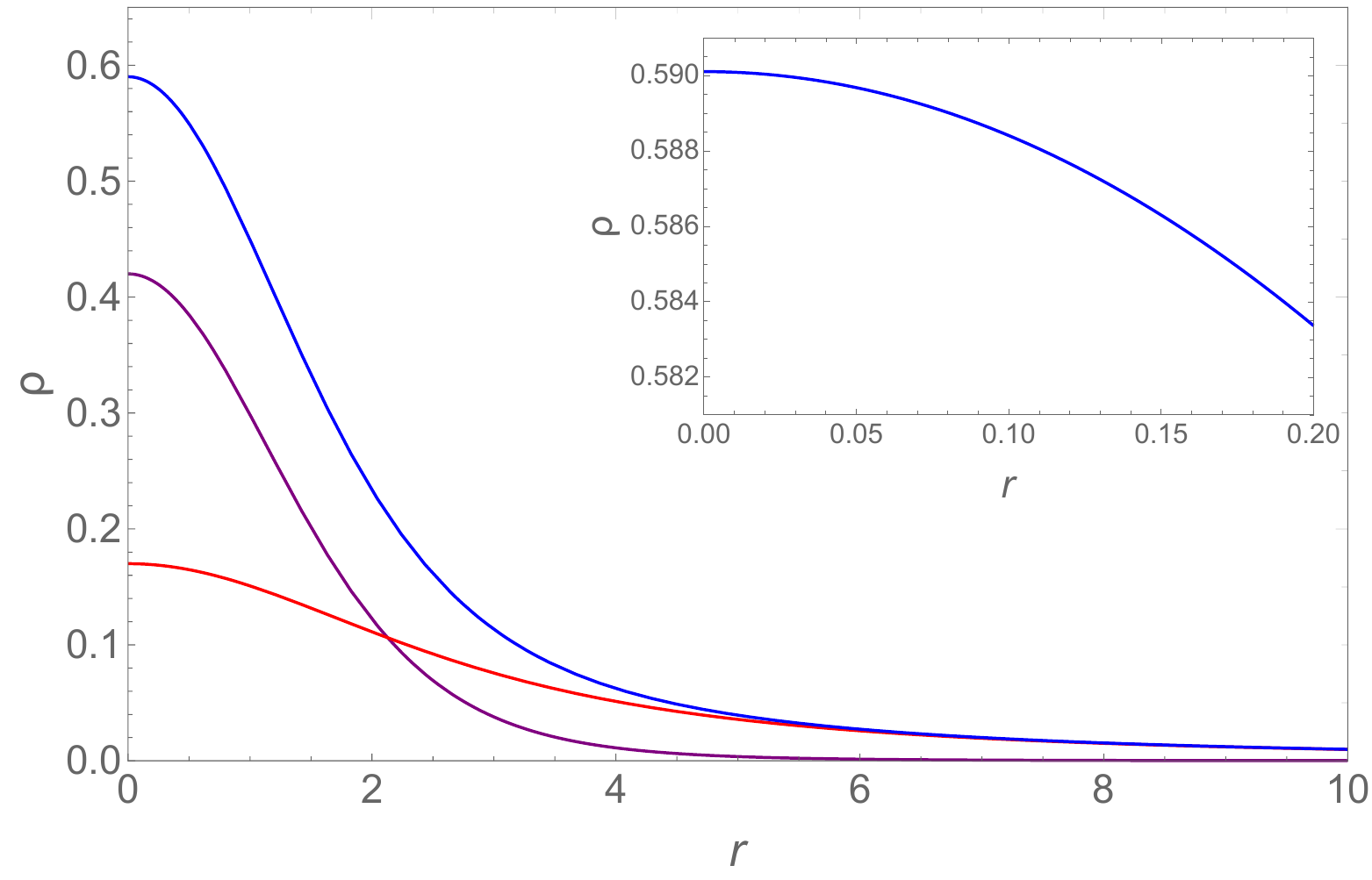}
    \caption{String energy density $\rho$ (blue) as a function of radius $r$ for $\lambda=1$, showing both the core contribution from massive modes (purple) and exterior massless modes (red) which dominate beyond $r\gtrsim 2$.  Inset is the energy density close to $r=0$ showing a flattened centre which allows small `zero-mode' excitations within the string core $\Delta r \approx {\cal O} (\delta/10)$.}
    \label{fig:StringEnergyDensity}
\end{figure}

\begin{figure}
    \centering
    \includegraphics[width=\linewidth]{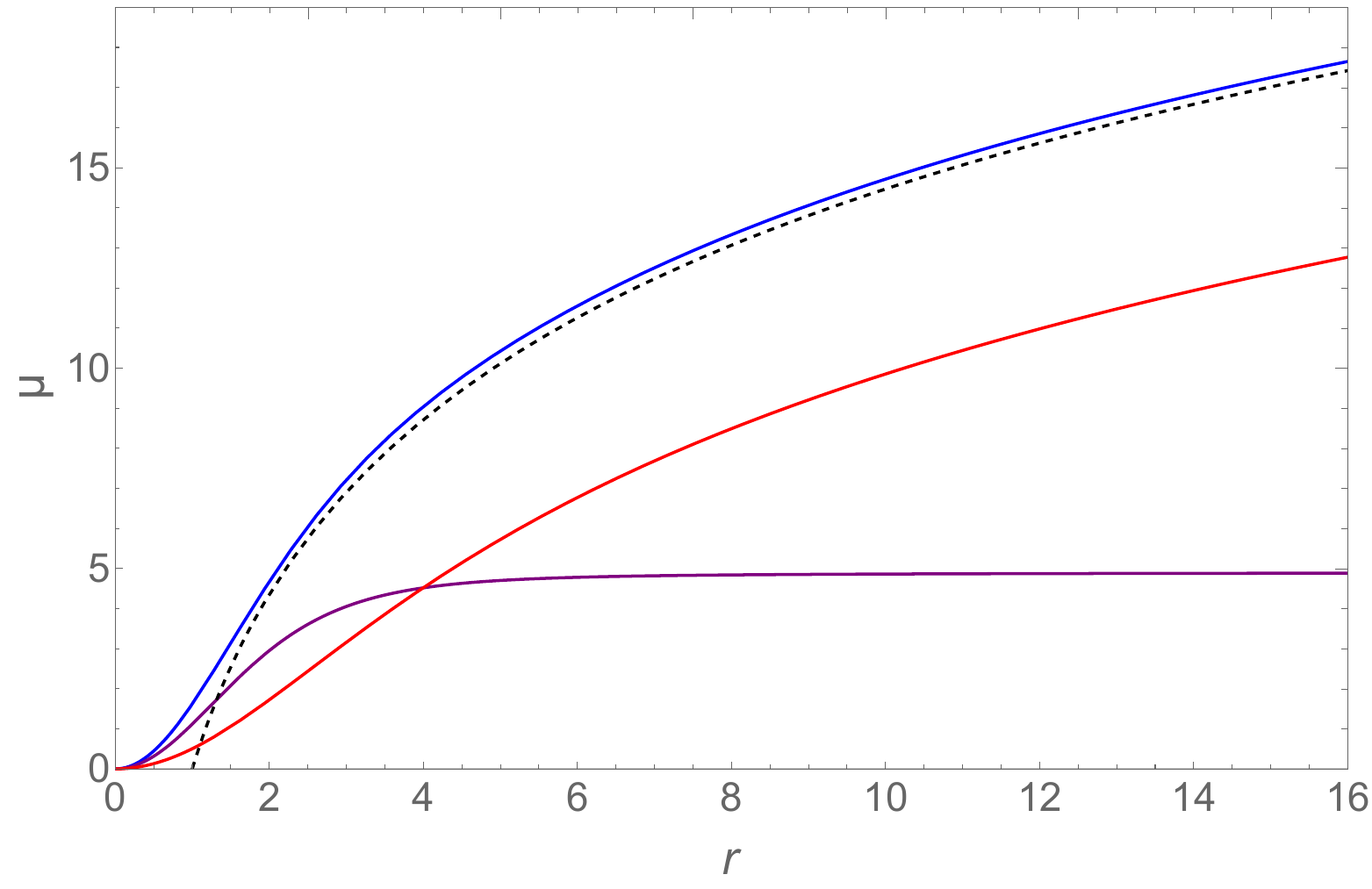}
    \caption{Integrated string energy per unit length $\mu(R)$ as a function of cutoff radius $r=R$ for $\lambda=1$, showing the total energy density (blue), the massive core (purple) and the long-range massless contribution (red) which renormalizes $\mu$.  Agreement with the simple logarthmic fit $\mu \approx 2\pi \ln(R/\delta)$ is shown for $R\gtrsim 2$ (dotted line).}
    \label{fig:MuIntegrated}
\end{figure}

The energy density $\rho(r) = T^{00}$ of the string in cylindrical coordinates can be calculated from the energy-momentum tensor 
\begin{equation}
T_{\mu\nu} = \partial_\mu\varphi\partial_\nu\varphi - g_{\mu\nu}\mathcal{L}\,
\end{equation}
and splits into the following contributions: 
\begin{equation}\label{radialenergydensity}
\rho(r) =  \left(\frac{d\phi}{dr}\right) ^2 + \frac{\lambda}{4} \left (\phi^2-1\right)^2 + \left(\frac{n_w\phi}{r}\right)^2\,,
\end{equation}
where $((1/r)|\partial\varphi/\partial\theta|)^2 = (n_w\phi/r)^2$ from the string cross-section \eqref{phi}.
The first two terms are the gradient and potential energies associated with deviations of the massive field from the vacuum, i.e. for $\phi \lesssim 1$. These provide the dominant contribution to the `local core' within a radial distance $r\lesssim2\delta$, as illustrated in Figure~\ref{fig:StringEnergyDensity}.  This massive contribution to the energy density converges rapidly to the vacuum as $1-r^{-2}$, and integrating out to $r\rightarrow \infty$ yields a core energy density $\mu_0 = 4.9$.  The third contribution is due to the `winding' of the long-range massless field about the string core, which generically dominates the overall energy density beyond $r\gtrsim 2\delta$ (also shown in Figures~\ref{fig:StringEnergyDensity}--\ref{fig:MuIntegrated}).  In principle, this massless contribution is logarithmically divergent,
\begin{equation}
\label{muintegrate}
\mu_\theta(R) \approx \int^{R}_{\delta}\left|\frac{1}{r}\frac{\partial\varphi}{\partial\theta}\right|^2 2\pi r dr  = 2\pi \eta^2\ln\left(R/\delta\right)\,.
\end{equation} 
However, in practice, it will be cut off at some radius $R$ associated with the curvature radius of the string, at which point the correlations implied by the \textit{ansatz} (\ref{phi}) will be washed out or cancelled.   For axion strings on cosmological scales, we expect $\mu_\theta \gg \mu_0$. In fact, (\ref{muintegrate}) provides an accurate estimate even on much smaller scales, with the total energy density
\begin{equation}
\label{muapprox}
\mu(R) = \mu_0 + \mu_\theta (R) \,\,\approx\,\, 2\pi\eta^2\ln\left(R/\delta\right)~~\hbox {(for $R\gtrsim 2$)}\,,
\end{equation} 
achieving better than 2\% accuracy for $R\gtrsim10$ and 0.1\% on cosmological scales (see Figure~\ref{fig:MuIntegrated}).

The internal structure of global strings is an important factor for numerical simulations, which inherently have a limited dynamic range. We can identify the positions of string cores in space ${\bf x}$ by locating the zeroes of the field where $\phi({\bf x})=0$. However, this might not represent the actual string centre of mass because of internal excitations within the string radius.  Inset in Figure~\ref{fig:StringEnergyDensity} is a zoomed view of the energy density around $r=0$ close to the string core, showing that it becomes  `flat' and varies by only 1\%  ($\rho_0-\rho(r)\lesssim 0.01$) within a radius $r\lesssim 0.2\delta$.  This means there is an internal `zero mode' allowing the string core to move short distances at little or no energy cost, without actually moving the bulk string.  We can expect such finite width effects to be present when studying small amplitude oscillations comparable to the string width $A\sim \delta$. 

\subsection{Massive and Massless Radiation Modes}\label{masslessmassive}

An oscillating global string will emit both massless (Goldstone) modes and massive (Higgs) radiation. Analytic expectations are very different for these two channels, so it is important to develop robust diagnostic tools to be able to numerically extract and analyse them separately.   This is a significant challenge because radiative modes must be separated from string self-fields, which, in principle, can also be long-range and time-varying.  We discuss further how propagating modes can be distinguished from self-fields in Section~\ref{radvsself}.

\begin{figure}
    \centering
    \includegraphics[width=0.9\linewidth]{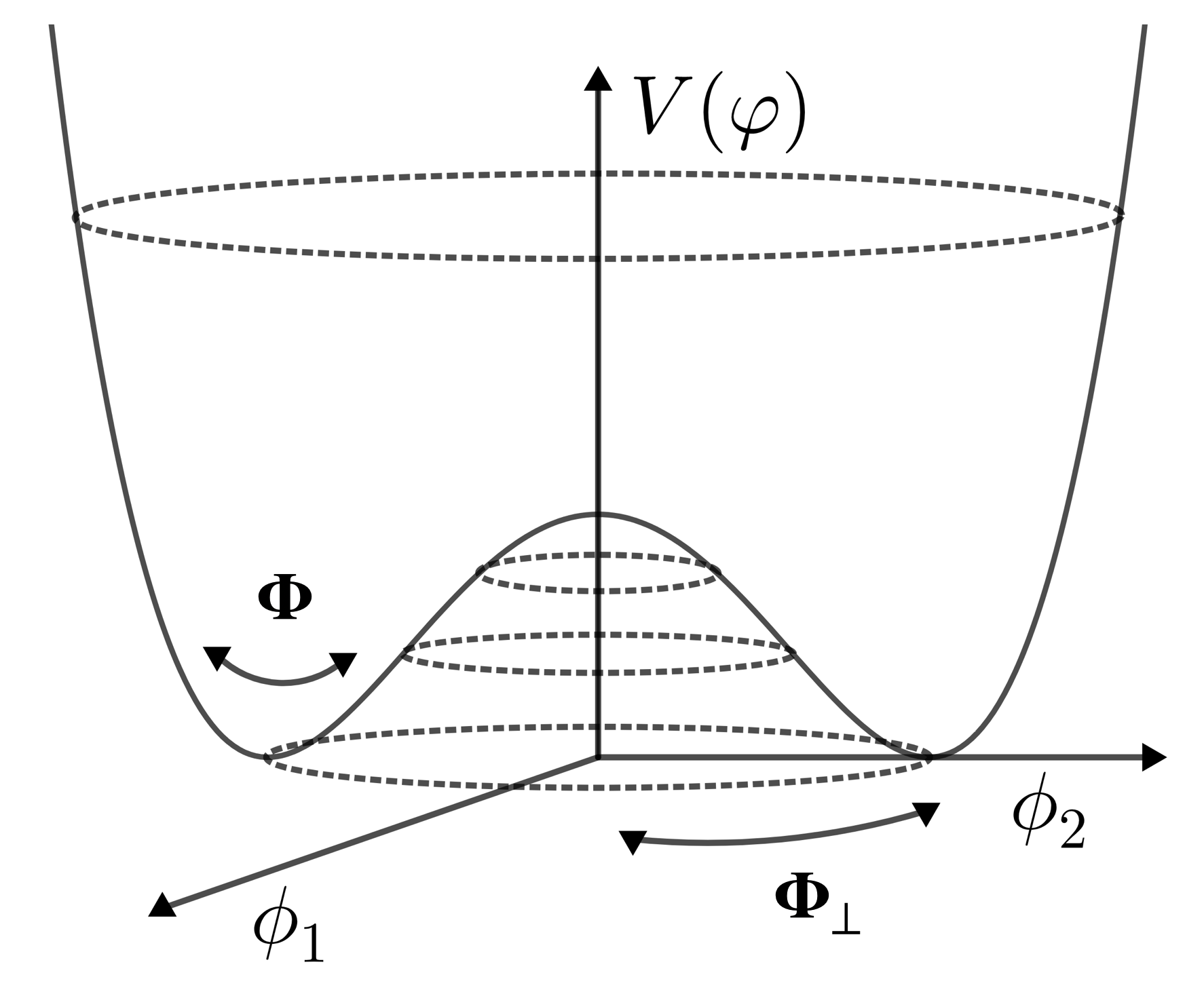}
    \caption{Diagram of symmetry breaking potential $V(\varphi)$, with orthogonal directions ${\mathbf{\Phi}}$ and
    ${\mathbf{\Phi}}_\perp$ indicated (see also (\ref{diagnosticrelations1}-\ref{diagnosticrelations2})).}
    \label{fig:mexicanhat}
\end{figure}

We first demonstrate the presence of massive and massless radiative modes around the broken symmetry vacuum state (\ref{vacuumbrokensym}) using the general form of the Argand representation 
\begin{equation} \label{Argand}
    \varphi(x^\mu) =  \phi(x^\mu)\,e^{i\,\vartheta(x^\mu)}\,,
\end{equation}
where both the magnitude $\phi (x^\mu)=|\varphi(x^\mu)|$ and the phase $\vartheta(x^\mu)$ are real scalar fields associated with the orthogonal excitations illustrated in Figure~\ref{fig:mexicanhat} (and we have set $n_w=1$).  The field equations (\ref{EL}) split into real and imaginary parts, respectively, as
\begin{align}\label{real}
    \frac{\partial^2\phi}{\partial t^2} -\nabla^2\phi &= \phi \left [\left(\frac{\partial\vartheta}{\partial t}\right)^2-(\nabla\vartheta)^2+{\frac{\lambda}{2}} (1-\phi^2)\right],\\
    \frac{\partial^2\vartheta}{\partial t^2} - \nabla^2\vartheta &= \frac{2}{\phi}\left(\frac{\partial\phi}{\partial t}\frac{\partial\vartheta}{\partial t} - \nabla\phi \nabla\vartheta\right) \,. \label{imag}
\end{align}
Assuming that $\vartheta$ is nearly constant far from any strings, (\ref{real}) becomes
\begin{equation}
    \frac{\partial^2\phi}{\partial t^2}-\nabla^2\phi - {\frac{\lambda}{2}}  \phi \left(1-\phi^2\right) = 0\,.  \label{massivediag}
\end{equation}
Expanding around the vacuum state $|\varphi|=\eta$ (where we have taken $\eta=1$) by setting $\phi = 1 + \chi$, we linearise to obtain the Klein-Gordon equation
\begin{equation}\label{KleinGordon}
    \frac{\partial^2\chi}{\partial t^2}-\nabla^2\chi + m_H^2\chi = 0\,,
\end{equation}
where $m_H = \sqrt\lambda\, \eta$.  Hence, in this limit, we deduce that $\chi$ acts like a free massive scalar field. On the other hand, if $\phi$ is nearly constant, the second equation (\ref{imag}) reduces to the wave equation 
\begin{equation}\label{masslessradiation}
    \frac{\partial^2\vartheta}{\partial t^2}-\nabla^2\vartheta = 0\,,
\end{equation}
where $\vartheta$ behaves as a massless scalar field.   Asymptotically far from any strings, it should therefore be a good approximation to decompose radiation into these distinct massive and massless modes.

In order to determine individual contributions to the radiative spectrum from each mode, it will be useful to measure and separate the components of the energy-momentum tensor, given by 
\begin{eqnarray}\label{energymomentum}
   T_{\mu\nu} = 2\partial _{(\mu} \bar\varphi \partial _{\nu)} \varphi - g_{\mu\nu} \left ( \partial _\sigma\bar\varphi\partial ^\sigma\varphi -  {\textstyle\frac{\lambda}{4}} (\bar \varphi \varphi -1)^2\right).
\end{eqnarray}
We can decompose the energy density $\rho = T^{00}$ into massive and massless modes using (\ref{Argand}) as follows:
\begin{align}\label{energydensity}
   T^{00}  &= \dot {\bar\varphi}\dot \varphi +\nabla_i {\bar\varphi}\nabla_i \varphi + {\textstyle\frac{\lambda}{4}} (\bar\varphi\varphi -1)^2 \\
&= \dot \phi ^2 + (\nabla\phi)^2   + \phi^2 \left( \dot \vartheta^2 + (\nabla \vartheta)^2 \right )+ {\textstyle\frac{\lambda}{4}} (\phi^2 -1)^2 \nonumber\\
&=  \dot \phi_1 ^2 + \dot \phi_2 ^2 + (\nabla\phi_1)^2  +  (\nabla\phi_2)^2  + {\textstyle\frac{\lambda}{4}} (\phi_1^2 +\phi_2^2 -1)^2 \nonumber\,,
\end{align}
where in the last line we have reintroduced the complex components (\ref{complex}), $\varphi = \phi_1+i \phi_2$. To make clear the momentum components of the fields in the $\theta$- and $r$- directions, we introduce  $\mathbf{\Phi}=(\phi_1,\phi_2)$. where $\hat{\mathbf{\Phi}} = {\mathbf{\Phi}} / |{\mathbf{\Phi}}|$ represents the radial direction in field space, and $\mathbf{\Phi}_\perp=(\phi_2,-\phi_1)$ which is orthogonal, as shown in Figure~\ref{fig:mexicanhat}. From this we note the relations
\begin{align}
    (\dot {\mathbf{\Phi}} \cdot\hat{\mathbf{\Phi}})^2 &= \dot\phi^2 = \frac{1}{\phi^2} \left(\phi_1\dot\phi_1 + \phi_2\dot\phi_2\right)^2    \label{diagnosticrelations1}  \,,\\
   (\dot {\mathbf{\Phi}}\cdot \hat {\mathbf{\Phi}}_\perp)^2 &= \phi^2\dot\theta^2 = \frac{1}{\phi ^2} \left(\phi_2\dot \phi_1 - \phi_1\dot\phi_2\right)^2 \,.
    \label{diagnosticrelations2}
\end{align}
 Extrapolating from (\ref{diagnosticrelations1}) and (\ref{diagnosticrelations2}), we can therefore deduce a direct numerical diagnostic for the distinct massive and massless components of $T_{\mu\nu}$ by defining the following real momenta and spatial gradients
\begin{align}\label{massivediagnostic}
 &{\Pi}_\phi \equiv \frac{\phi_1\dot\phi_1 + \phi_2\dot\phi_2}{\phi}, ~~ {\cal D}_i \phi \equiv  \frac{\phi_1\nabla_i\phi_1 + \phi_2\nabla_i\phi_2}{\phi}, \\
&{\Pi}_\vartheta \equiv \frac{\phi_1\dot\phi_2 - \phi_2\dot\phi_1}{\phi}, ~~   {\cal D}_i \vartheta \equiv  \frac{\phi_1\nabla_i\phi_2 - \phi_2\nabla_i\phi_1}{\phi}.\label{masslessdiagnostic}
\end{align}
 We can use these to express the energy density (\ref{energydensity}) in terms of massive and massless components in the following form:
 \begin{eqnarray}\label{energymomdiag}
   T^{00} =  {\Pi}_\phi^2 + ( {\cal D} \phi)^2 + {\Pi}_\vartheta^2 + ( {\cal D} \vartheta)^2 +  {\textstyle\frac{\lambda}{4}} (\phi^2 -1)^2.
\end{eqnarray}
Furthermore, the relations  (\ref{massivediagnostic}) and (\ref{masslessdiagnostic}) allow us explicitly to split the momentum component $T^{0i}$ of the stress tensor into massive and massless components, given by
\begin{eqnarray}\label{momentumdiag}
   P_i\equiv T^{0i} =  2({\Pi}_\phi {\cal D}_i \phi + {\Pi}_\vartheta{\cal D}_i \vartheta) \,.
\end{eqnarray}
For our massive and massless scalar radiation fields, the two quantities in (\ref{momentumdiag}) are equivalents of the electromagnetic Poynting vector describing radiation energy fluxes.   Choosing an outgoing radial direction, we can integrate the two components of ${\bf P}\cdot \hat{\bf r}$ on a distant surface to determine the energy flow out of the enclosed volume for each type of radiation. 

In this paper, we want primarily to analyse radiative modes that propagate outwards from the string. The massless radiation field and string self-field can be of the same order of magnitude in the energy density $\rho$, making them difficult to distinguish. However, the propagating contribution can be effectively separated out using the spatial diagnostic ${\bf{\cal D}}\vartheta \cdot \hat {\bf r}$. As we shall see in Section \ref{Massless}, time variations of the self-field can be described asymptotically as non-propagating solutions which are low harmonics of the  fundamental frequency of the string. These have weak spatial gradients in the radial direction, so ${\bf{\cal D}}\vartheta \cdot \hat {\bf r}$ is small.  This diagnostic is a very useful quantity for visualisation, producing much cleaner massless radiation signals than the momentum ${\Pi}_\vartheta$ which is more strongly contaminated by the self-field. Massive radiation proves more complex, with both $\Pi_\phi$ and ${\bf{\cal D}}\phi \cdot \hat {\bf r}$ required to provide a full picture of the propagating modes. The time evolution of the massless diagnostics will be analysed in subsequent sections using Fourier transforms to determine the total spectral composition of the outgoing radiation.

\subsection{Separation of scales and current discrepancies}\label{ScalesSep}

As outlined in the Introduction, two string simulation methods have been predominantly used to investigate cosmic and axion string dynamics; 1D vortex-lines solving Nambu-Goto equations, offering a huge dynamic range, or 3D solutions of the full three-dimensional field theory, including radiation but with a limited dynamic range.  These different approximations appear to produce different outcomes, which are interpreted by some researchers as contradictory. This has been an outstanding concern in the literature on string network simulations for many years, so renewed effort is needed to establish whether these approaches converge on cosmological scales. 

An important consequence of these alternative approaches is quantitative uncertainty about the amplitude and spectrum of string network decay products. Topological strings radiate primarily into the lowest mass channels available, which include axions (or Goldstone bosons) for axion (or global) strings and gravitational waves for `local' cosmic strings (arising from a broken gauged symmetry).  Despite the apparent simplicity of global axion strings, the resulting spectrum of axion radiation in particular has proved controversial to characterise due to the limited dynamic range of numerical simulations. From (\ref{muapprox}), a global axion string has the logarithmically divergent linear energy density
$
\mu = 2\pi f_a^2\mathrm{ln}(R/\delta)\,,
$
where $f_{\rm a}$ is the Peccei-Quinn energy-breaking scale.   For a typical dark matter axion model with $f_{\rm a} \sim 10^{11}\,$GeV, the string width $\delta\sim 10^{-23}\,$m, whereas $R \sim 1/200\,$MeV$\,\sim 1\,$m at the QCD scale. The separation of scales in a cosmological context is therefore given approximately by
  \begin{equation}\label{ln70}
\ln (R/\delta) \sim 70\,.
 \end{equation}
In contrast, field theory numerical simulations have difficulty probing a dynamic range larger than $\ln (R/\delta) \sim 8$, not taking into account relativistic effects. This means that numerical axion strings generically have an order of magnitude stronger relative coupling to massless radiative modes than their cosmic counterparts.   For this reason, rather than extrapolating results from strongly coupled numerical simulations of axion strings, most authors have argued that the renormalisation offered by the logarithmic term (\ref{muapprox}) means that cosmological global strings behave qualitatively more like (local) Nambu-Goto strings \cite{Davis1986,Vilenkin:1986ku,Battye1993}.  On this basis, semianalytic approaches are used to estimate network radiation into axions \cite{Battye:1994au,Wantz:2009it}. 

For global cosmic strings near the GUT scale, the difference between typical length scales is even greater, given by
\begin{equation}\label{ln100}
\ln (R/\delta) \sim 100\,,
 \end{equation}
with less than 1\% of the energy per unit length localised in the string core $\mu_0/\mu$ (see (\ref{muapprox})).  
Under these circumstances, one would expect the Nambu-Goto approach (coupled to a massless scalar field) to be a good approximation.  However, despite the much larger dynamic range available to Nambu-Goto simulations in comparison to field theory simulations, it remains challenging to establish that the large-scale properties of networks converge to the predicted scale-invariant behaviour. Even more challenging to determine are the continuously evolving, fractal-like, small-scale features, including the loop production function which is key in analytic Nambu-Goto models (for example, see \cite{Blanco-Pillado2013}). 
(For axion strings, the presence of much stronger radiative effects should influence and probably stabilise small-scale network properties.) Moreover, there is no clear prescription for including radiative backreaction in Nambu-Goto simulations (although see \cite{Battye1995}). In the context of gravitational wave predictions, disagreements in the literature have meant that the constraints on local cosmic strings recently published by LIGO  consider three separate Nambu-Goto models \cite{Abbott2018}.

In contrast, field theory simulations evolve the `real' string equations of motion given by equation (\ref{EL}), allowing the full dynamics of the internal degrees of freedom to be captured. However, as discussed previously, with current computational resources it is impossible to simulate string networks using fixed grid simulations with sufficient dynamic range to achieve convergent behaviour. It is usually necessary to adopt the so-called `fat string' approach, growing the string width to match the comoving grid resolution \cite{PressRydenSpergel, Vincent:1997cx,Moore:2001px,Moore:2017ond,Gorghetto:2018myk,Hindmarsh:2019csc}, effectively lowering the particle mass $m_H$ and keeping light massive radiative channels competitive with massless modes. There have been sophisticated attempts to extrapolate from field theory simulations \cite{Gorghetto:2018myk,Kawasaki:2018bzv,Hindmarsh:2019csc}, but generally asymptotic scaling regimes differ on large scales by a factor of two from Nambu-Goto strings and small-scale features are strongly affected by radiative effects around the comoving string width $\delta$.  Innovations adding more gauged fields \cite{Klaer2017} have enabled global axion string simulations with larger tensions comparable to (\ref{ln70}) to be performed. However, this approach still uses the comoving width algorithm, so there remain artificially more massive radiation channels available. 

These alternative approaches yield different predictions for radiation rates from global cosmic strings, in the case of axion strings yielding incompatible dark matter axion mass predictions and an uncertain guide for axion searches \cite{Marsh:2015xka}. It is hence necessary to introduce new high resolution numerical  techniques to accurately resolve these differences, concentrating computational power where it is needed near the radiating string core. 

\section{Adaptive Mesh Refinement \& Simulation Setup}\label{AMR}

\subsection{AMR and GRChombo}

In order to accurately numerically evolve non-linear physical systems, it is vital that simulations are able to resolve features that emerge on different length scales. To resolve small-scale features requires a simulation box with a sufficiently fine mesh, whereas to capture macroscopic effects, we require a box size that is sufficiently large. Traditional `fixed grid' numerical approaches will often be unable to satisfy this requirement within the constraints of limited computational resources. 

One method that can be used to address this issue is to adapt the resolution of the numerical grid to the scale of the features of interest. This technique, known as `mesh refinement,' allows computational power to be concentrated in regions where the most refinement is needed. This can increase the size and/or precision of simulations that can be performed using a given amount of computational resources \cite{Kunesch2018}. Mesh refinement is widely used, for example, in numerical relativity simulations of black hole collisions (e.g. \cite{Cactus7}). More specifically, adaptive mesh refinement (AMR) refers to a particularly flexible approach which allows the level of refinement and the position, size and shape of the refined regions to be calculated and adapted as the simulation progresses, using a physically-motivated `tagging criterion'. This ensures that regridding is only performed in areas where it is physically required, without the need for a lot of prior knowledge about the behaviour of the system.

\begin{figure}
    \centering
    \includegraphics[width = 0.48\textwidth]{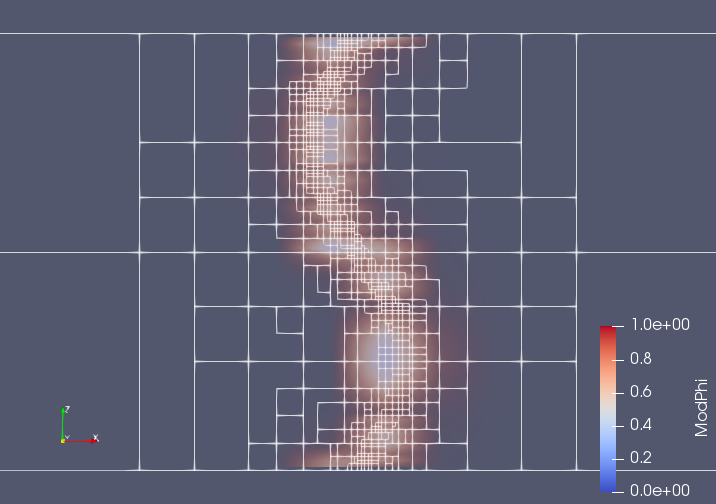}
    \caption{Snapshot of an AMR simulation of a global string using GRChombo. This figure shows $\phi$ for a sinusoidally displaced string with the outlines of AMR boxes for different refinement levels. Smaller boxes concentrated towards the centre of the string indicate areas where a finer mesh has been used.}
    \label{AMRstring}
\end{figure}

In this paper, we use GRChombo \cite{Clough2015}, an open-source finite difference AMR code that allows for refinement of the simulation grid `on-the-fly'. Originally designed for numerical relativity, we use GRChombo to evolve the global cosmic string field equations without gravity, whilst leaving the door open for future analysis in full general relativity. GRChombo  uses an AMR implementation based on the Berger-Rigoutsos mesh refinement algorithm \cite{Berger1991}. First order partial differential equations (PDEs) are evolved using a fourth-order Runge-Kutta (RK4) method, implemented first on a coarse base grid. The user then imposes a `tagging criterion' that determines where to refine the mesh to higher precision. In our case of a complex scalar field, this is given by
\begin{equation}
    \Delta x \sqrt{(\nabla\phi_1)^2 + (\nabla\phi_2)^2} > |\phi_{threshold}|\,,
\end{equation}
where $\Delta x$ is the grid spacing and $|\phi_{threshold}|$ is a custom threshold input by the user. If this criterion is met, the simulation will refine that area of the numerical grid. GRChombo outputs hdf5 files as it runs which can be easily visualised using software such as Paraview (or the output can be viewed directly using in-situ visualisation). An example of this is shown in Figure \ref{AMRstring}, highlighting the hierarchy of refinement levels concentrated around the string core. 

GRChombo heavily exploits both MPI and OpenMP parallelism, splitting the simulation box into smaller `AMR boxes' and using load-balancing to distribute these over multiple processors, as well as to execute multiple threads within each. This allows work to be spread evenly between processors, so that once coarser areas of the simulation have finished running, areas with higher refinement can be allocated to the idle processor. This means that less computational time is wasted and resources are used more efficiently. The most up-to-date profiling data for GRChombo is given here \cite{Kunesch2018}, along with further information about the code. The code has already been used in a wide range of applications \cite{Clough2018, Helfer2019, Clough2017, Lim2017, Helfer2017, Figueras2016, Widdicombe2018, Figueras2017, Helfer2018, Clough2016, Alexandre2018, Joana2020, Bamber2020}, from black hole-axion star collisions to Abelian-Higgs string loop collapse. 

\subsection{Dynamic Range}

As discussed in Section \ref{ScalesSep}, it is vital for the accuracy of string simulations that the string core is appropriately resolved. In this paper, we apply AMR to global strings in order to investigate whether the approximations commonly used for the string width give accurate results for string evolution and radiation. Here, we discuss the separation of scales that these current simulations reach, as well as predictions for future simulations.

In this investigation, we use a base-level box size of $256\times256\times32$ and probe string widths spanning over an order of magnitude, determined by the parameter $\lambda$ in the range $1\le \lambda \le 100$. (In fact, we explored a wider range $0.3\le \lambda \le 300$; although the extremes are not presented in this paper, a detailed investigation of $\lambda < 1$ in particular is presented in upcoming work \cite{Drew2020}.) Taking the radius of curvature $R$ to be half of the box length and $\delta \sim 1/\sqrt{100}$ for the narrowest string $\lambda=100$, this means that, for these relatively small simulations, we are probing the ratio
\begin{equation}
    \ln{(R/\delta)} \sim 7\mathrm{-}8\,.
\end{equation}
This is similar to the maximum value that can currently be reached using fixed grid methods. The maximum level of refinement in this simulation with $\lambda=100$ uses $\Delta x = 1/128$, so the string core $\delta$ is resolved by $\sim 10$ gridpoints. This is in contrast to many field theory simulations in the literature, where the string core is resolved by one or two gridpoints only. If we were to instead use this criterion, this level of refinement would be sufficient to resolve $\lambda=10,000$, corresponding to the ratio $\ln{(R/\delta)} \sim 9\mathrm{-}10$.\footnote{In some of our simulations, we found that $\Delta x \sim 1/32$, i.e. approximately three gridpoints resolving the string core, was insufficient to accurately capture the decay of the $\lambda=100$ string. We would expect this effect to be exacerbated when considering larger amplitudes with more relativistic motion, in part, because relativistic effects (i.e. length contraction) necessitate higher resolution. This indicates that using $\mathcal{O}(1)$ gridpoints to resolve the string core may be insufficient for network simulations in which large amplitude configurations appear to be common. Further investigation of this effect is required.}

As with fixed grid methods, the limiting factor for AMR simulations is computational power. We can estimate a loose lower bound on the memory required to resolve a string over a certain number of refinement levels $l$ by considering a cubic simulation box with refinement concentrated along a single straight string which spans the box length $L$, with the string width spanned by one cubic AMR box of each level. The total number of datapoints up to a given AMR refinement level $l_{\rm max}$, using a refinement ratio $r=2$, is given approximately by $(L/\Delta x_{0})^3 \,2^{l_{\rm max}+1}$. Here, we have summed over the levels $0 < l < l_{\rm max}$ which each use a number of AMR boxes $2^l$. Each box has $(L/\Delta x_{0})^3$ datapoints. For a cubic simulation box, this concentration of power at the centre saves a factor of $\sim 2^{2l_{\rm max}-1}$ in memory, e.g. $\sim 2^{13} \sim 10^4$ for one string with $l_{\rm max} = 7$. This saving becomes even more significant for non-cubic boxes, such as the ones used in this paper, with approximately another factor of 2 for each doubling of the $x$- or $y$- dimensions. Therefore, for a given amount of RAM, a fixed grid simulation will be able to reach an equivalent maximum level of resolution $l_{\rm max, \,fixed} = (l_{\rm max, \,AMR}+1)/3\,,$
corresponding to a factor difference in resolution of $\sim 2^{\frac{2}{3}l_{\rm max,\,AMR}}$. Scaling $\delta$ appropriately, this corresponds to a logarithm of $\sim \ln(R/(2^{\frac{2}{3}l_{\rm max,\,AMR}}\delta))$. For our example of a single string with $l_{\rm max,\,AMR}=7$, this corresponds to a $\ln(R/\delta)$ approximately $3$$-$$4$ lower than the AMR simulation. Using these crude estimates and taking a recent literature value of $\left.\ln(R/\delta)\right|_{\rm fixed} = 8$ from \cite{Gorghetto:2021} for fixed grid network simulations, we could similarly expect AMR network simulations probing higher logarithms by $\cal{O}$$(2$$-$$3)$ to be possible using the same resources, with further potential savings when using a higher refinement ratio.

However, although there are impressive savings in memory, unfortunately AMR is not a straightforward panacea. The regridding and reconstruction of the finer AMR levels incurs an overhead cost that is not present for fixed grid simulations. This offsets some of the memory savings and, at a certain level of resolution, will lead to AMR simulations becoming unfeasible. Nevertheless, AMR provides an important step towards probing higher string energy scales.

\subsection{Initial Conditions}

We simulate the evolution of a sinusoidally displaced global string for a range of $\lambda$. Initial conditions are obtained first by numerically solving the static field equation (\ref{radialeqn}) to obtain the radial profile $\phi(r)$, as discussed in Section \ref{globalstrings}. We use this $\phi(r)$ to set the initial values
\begin{equation}
    \phi_1 = \phi\cos{(n_w\theta)}, \hspace{1cm} \phi_2 = \phi\sin{(n_w\theta)}\,,
\end{equation}
where $n_w$ is the string winding number.

These calculations provide us with the initial data for $\phi_1$ and $\phi_2$ in 2D, but this must be extended in the $z$-direction to create a 3D string. In the present study, we require an $n_w=1$ string that is sinusoidally displaced, so we create a crude initial approximation of the initial conditions by manually displacing the radial profile in the $x$-direction from $x=0$ as a function of the $z$-coordinate, using
\begin{equation}\label{sineapprox}
{\bf{X}} = \left(A\sin\left(\Omega_z z\right), 0, z\right)\,,  
\end{equation}
where $A$ is the initial amplitude, $\Omega_z=2\pi/L$ is the fundamental frequency (at small amplitude)  and $L=32$ is the approximate wavelength of the string, equivalent to the $z$-dimension of the box, as demonstrated in Figure \ref{AMRstring}. 

In order to have a general definition for large string amplitudes, we define the relative amplitude $\varepsilon$ as
\begin{equation}\label{relamp}
\varepsilon ~\equiv~ \frac{2\pi A}{T}   ~\approx~  \frac{2\pi A}{L}~~ \text{for} ~A\ll L\,,
\end{equation}
which characterises the relationship between the amplitude $A$, the spatial periodicity $L$ and the actual oscillation period $T$.  Note that the time period of oscillations $T$ is related to the invariant length of the string which is longer than the spatial periodicity defined by the box sidelength $L$.  Keeping to small amplitudes $A\ll L$, then we have $T\approx L$. However, the larger $\varepsilon$ becomes, the further away the configuration \eqref{sineapprox} is from being an appropriate \textit{ansatz} for the string initial conditions. We hence need to relax this initial sinusoidal configuration to obtain lower energy initial data. 

To achieve this we employed gradient flow methods to create initial conditions using dissipative evolution:
\begin{equation}
    \frac{\partial\varphi}{\partial t} -
    \nabla^2\varphi + \frac{\lambda}{2}\varphi(\varphi\bar{\varphi} - \eta^2) = 0\,,
\end{equation}
starting with a considerably larger initial $A$ than the target amplitude $A_0$.  It was generally found to be sufficient to choose $A$ about 50\% larger than $A_0$ to obtain reproducible results in which the long-range fields were sufficiently relaxed.  

\subsection{Diagnostic Tools}\label{diagnostictools}

\subsubsection{Radiation Cylinder}
\begin{figure}
    \centering
    \includegraphics[width=\linewidth]{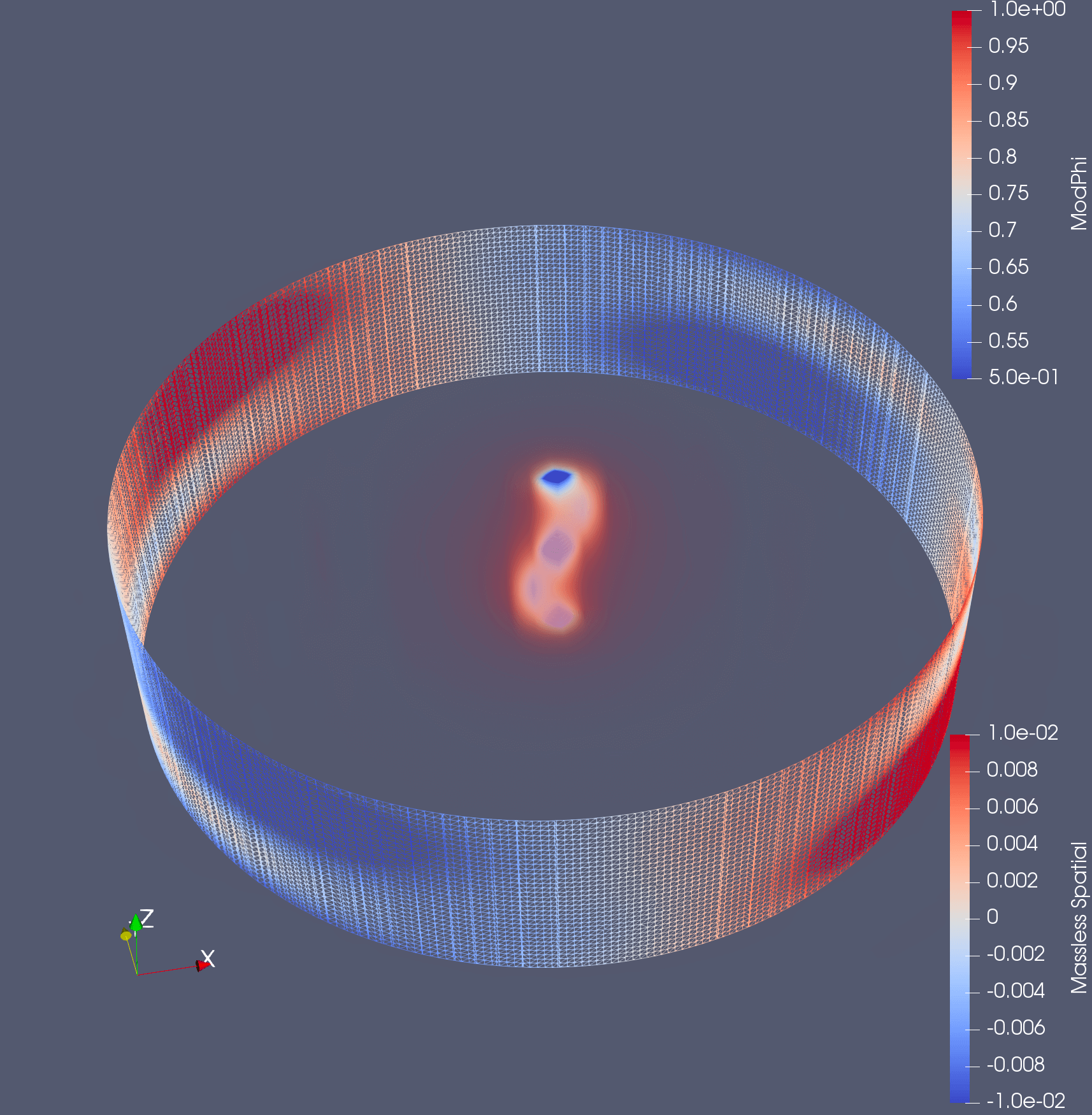}
    \caption{Extraction of the `spatial' massless radiation diagnostic ${\cal{D}}\vartheta \cdot \hat {\bf r}$ (bottom legend) on a cylinder at $R=64$. The string in the centre is depicted by $|\varphi| = \phi$ (top legend). We see a dominant quadrupole signal on the cylinder surface.}
    \label{PlaceholderCylinder}
\end{figure}

In order to extract the radiation emitted from the oscillating strings, we construct an analysis cylinder centred on the string core. We choose a radius of $R=64$, a distance far enough from the string core to minimise the effect of the self-field, but far enough from the boundaries to allow extended analysis before any radiation reflections can affect the central region. This diagnostic cylinder allows us to choose a field to sample at that radius and to extract the data on the cylinder. As outlined in Section \ref{masslessmassive}, in our analysis, we use the diagnostic $\cal{D}\vartheta \cdot \hat {\bf r}$ as defined by (\ref{masslessdiagnostic}) to analyse the massless radiation. An example of this setup is shown in Figure \ref{PlaceholderCylinder}. This is a similar technique to that used by the LIGO and Virgo Collaborations in their analysis of gravitational waveforms from binary black holes, although this instead uses spherical extraction due to the different overall symmetry.

\begin{figure}
    \centering
    \includegraphics[width=0.4\textwidth]{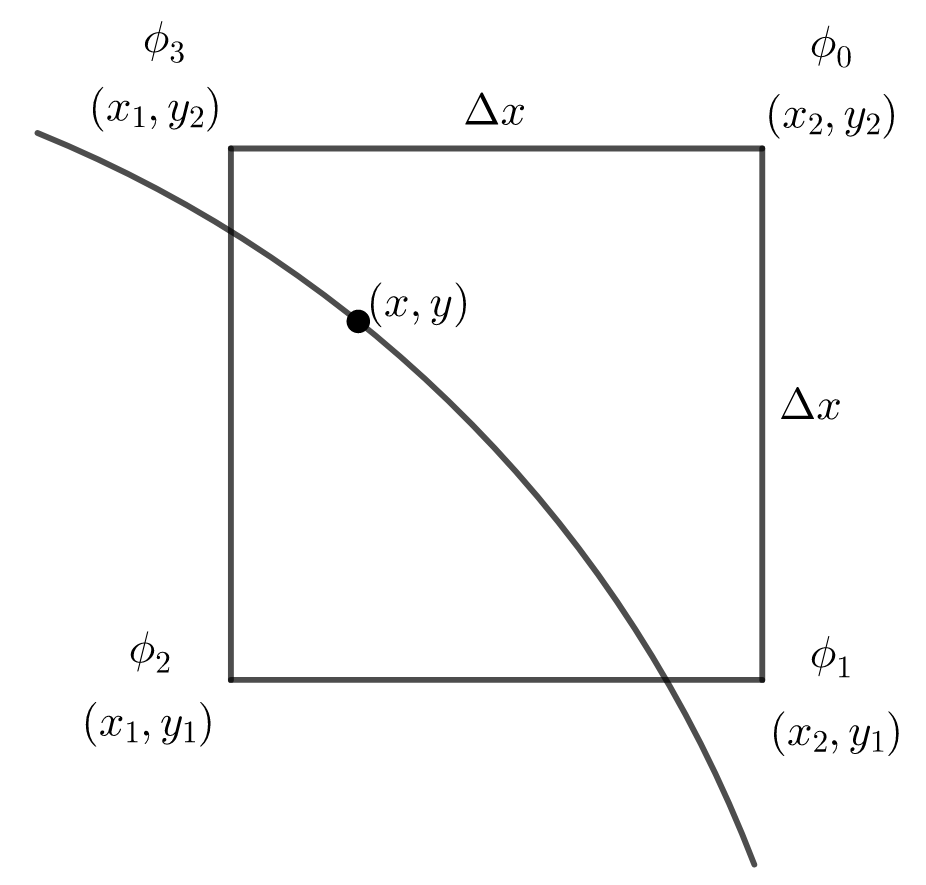}
    \caption{Diagram of a grid cell to demonstrate two diagnostic tools: \textit{1)} interpolation of radiation diagnostics onto a cylinder and \textit{2)} calculation of the position of the string core. For \textit{1)}, the arc represents a section of the cylinder on which radiation is analysed, where $(x,y)$ is the point onto which we interpolate the field values $\phi_i$ and $(x_i, y_j)$ are the coordinates of the corners of the cell. The value of the field $\phi$ at $(x,y)$ is calculated using equation (\ref{bilinear}). For \textit{2)}, the labels $\phi_i$ give the values of $\phi$ at the corners of a cell inside which has been detected a point of integer winding. We can substitute these values into (\ref{quadraticcoeffs}) to calculate the position of the string core.}
    \label{square}
\end{figure}

As the cylinder is defined on a Cartesian grid, it is necessary to interpolate values from the `nearest-neighbour' grid points to get an accurate value for the field on the surface itself. We first choose the number of points on the circumference of the cylinder to sample (here 256), and calculate their $(x,y,z)$ coordinates. We then use bilinear interpolation to determine the accurate value of the radiation fields at these points on the cylinder. The field value $\phi(x,y)$ is given by:

\begin{align}\label{bilinear}
    \phi(x,y) \approx
    \frac{1}{(x_2 - x_1)(y_2 - y_1)}
    \left[\phi(Q_{11})(x_2 - x)(y_2 - y)\right. \nonumber \\ + \phi(Q_{21})(x - x_1)(y_2 - y) + \phi(Q_{12})(x_2 - x)(y - y_1) \nonumber \\ \left.+ \phi(Q_{22})(x - x_1)(y - y_1)\right]
\end{align}
where $Q_{11}=(x_1, y_1),$ $Q_{12}=(x_1, y_2),$ $Q_{21}=(x_2, y_1),$ $Q_{22}=(x_2, y_2)$ and the coordinates $x_{1,2}$ and $y_{1,2}$ are defined as in Figure \ref{square}. From this, we obtain a $256\times32$ array of points (where 32 comes from the $z$-dimension of the box) for each timestep for both diagnostics $\Pi_\phi$ and $\cal{D}\vartheta \cdot \hat {\bf r}$, on which we can perform a 2D FFT to determine the Fourier decomposition.

\subsubsection{String Core Position}\label{stringcoreposition}
In addition to Fourier mode extraction, we also need to track the exact position of the string core to determine the rate of energy loss. We use a similar technique to the interpolation onto the diagnostic cylinder, but in reverse, scanning the domain to detect grid cells in which there is a non-zero winding. As we know that $\phi = 0$ at the string core, we can use the values of the complex scalar fields at the corners to fit a quadratic
\begin{equation}\label{quadratic}
    ax^2 + bx + c = 0
\end{equation}
to the grid cell shown in Figure \ref{square} and calculate the position of the zero. We simultaneously solve the system of equations
\begin{align}
    a(x_1 + iy_1)^2 + b(x_1 + iy_1) + c &= \phi_2 \nonumber \\
    a(x_1 + iy_2)^2 + b(x_1 + iy_2) + c &= \phi_3 \nonumber \\
    a(x_2 + iy_1)^2 + b(x_2 + iy_1) + c &= \phi_1 \nonumber \\
    a(x_2 + iy_2)^2 + b(x_2 + iy_2) + c &= \phi_0 \,
\end{align}
where $\phi_{0,1,2,3}$ are defined by Figure \ref{square} as the grid points at the corners of the relevant cell and $a, b$ and $c$ are constants. We obtain the coefficients
\begin{align}\label{quadraticcoeffs}
    a &= \frac{-i}{8}(\phi_0 + \phi_2 - \phi_1 - \phi_3) \nonumber \\
    b &= \frac{i-1}{8}(\phi_0 - \phi_2 + i( \phi_3 - \phi_1)) \nonumber \\
    c &= \frac{1}{4}(\phi_0 + \phi_2 + \phi_1 + \phi_3)\,.
\end{align}
for equation (\ref{quadratic}), which can be solved using the quadratic formula. The smallest root provides us with a fractional correction to the $x$-coordinate $x_{correct}$, such that
\begin{equation}\label{xcore}
x_{core} \approx x_{centre} + x_{correct}\frac{\Delta x}{2}\,,
\end{equation}
where $x_{core}$ is the true position of the string core and $x_{centre}$ is the $x$-coordinate of the centre of the grid cell. 
We use this to calculate the $x$-coordinate of the string core within the cell to second-order accuracy.

\subsection{Simulation Setup, AMR Parameters and Convergence Testing}\label{SimulationSetup}

In Section \ref{Massless}, we present, for the first time, simulations of global strings with string widths spanning over an order of magnitude, determined by the parameter $\lambda$ in the range $1\le \lambda \le 100$. Fixing the spatial periodicity of the strings at $L=32$ and energy scale $\eta =1$, we survey several perturbation amplitudes in the range $1\le A_0 \le 8$ (or relative amplitudes $0.20 \lesssim \varepsilon_0 \lesssim 1.0$) with initial conditions obtained using dissipative evolution. Using the diagnostic tools described previously, we analyse the propagating radiation modes generated by the strings, as well as the detailed string trajectory as its oscillation energy decays, directly comparing with the analytic predictions for massless modes. Over one hundred high resolution simulations are performed, using up to six levels of grid refinement.

All production simulations are carried out using a coarse simulation box size of $256\times256\times32$ ($N_1 \times N_2 \times N_3$) with periodic boundary conditions in the $z$-direction and Sommerfeld (outgoing radiation) boundary conditions in the $x$- and $y$- directions. A base grid of resolution $\Delta x_0 = 1$ is used with a base timestep $\Delta t_0 = \Delta x_0 / 4$. We choose a regridding threshold $|\phi_{threshold}| = 0.25$ for simulations with $\lambda < 10$ and $|\phi_{threshold}| = 0.1$ for those with $\lambda \geq 10$. These values were judged sufficient to capture the dynamics accurately by comparing convergence using different thresholds.

\begin{table}[t]
    \centering
    \caption{Grid parameters for the evolution stage of the convergence tests for the massless diagnostic $\mathcal{D} \theta$ and the amplitude $A$. We perform two variations of tests: \textit{i)} $l_{\rm max}$ is changed and the base grid resolution $\Delta x_0$ remains constant and \textit{ii)} $l_{\rm max}$ remains constant and $\Delta x_0$ changes. The base grid box resolution is given by $N_1 \times N_2 \times N_3$, with $(l_{\max}+1)$ total refinement levels including the coarsest base level, and grid spacings on the coarsest level given by $\Delta x_{l_{\rm max}}$. The length of the longest box side is given by $L_{\rm max}$. The grid parameters for the corresponding damping stages are identical, except that $l_{\max} = 1$.
    }
    \begin{ruledtabular}
    \begin{tabular}{c|ccccc}
        Test & $N_1 \times N_2 \times N_3$ & $l_{\max}$ & $L_{\max}$ & $\Delta x_0$ & $\Delta x_{l_{\rm max}}$  \\
        \hline
        \textit{i)} & $256\times 256\times 32$ & 0 & 256 & 1 & 1 \\
         & $256\times 256\times 32$ & 1 & 256 & 1 & 0.5 \\
         & $256\times 256\times 32$ & 2 & 256 & 1 & 0.25 \\
        & $256\times 256\times 32$ & 3 & 256 & 1 & 0.125 \\
        \hline
        \textit{ii)} & $64\times 64\times 8$ & 3 & 256 & 4 & 0.5 \\
         & $128\times 128\times 16$ & 3 & 256 & 2 & 0.25 \\
         & $256\times 256\times 32$ & 3 & 256 & 1 & 0.125 \\
    \end{tabular}
    \end{ruledtabular}
    \label{convergence_params}
\end{table}

\subsubsection{Convergence Testing}

We establish convergence of our simulations by measuring the massless radiation $\mathcal{D}\theta$ and the oscillation amplitude of a $\lambda=10$ string with $A_0=4$ for the grid configurations presented in Table \ref{convergence_params}. We obtain the initial conditions by dissipative evolution as in our production simulations, using the corresponding grid parameters to those in Table \ref{convergence_params} but with maximum refinement level $l_{\rm max}=1$, damping as close as possible to $A_0=4$.

\begin{figure}
    \includegraphics[width=0.5\textwidth]{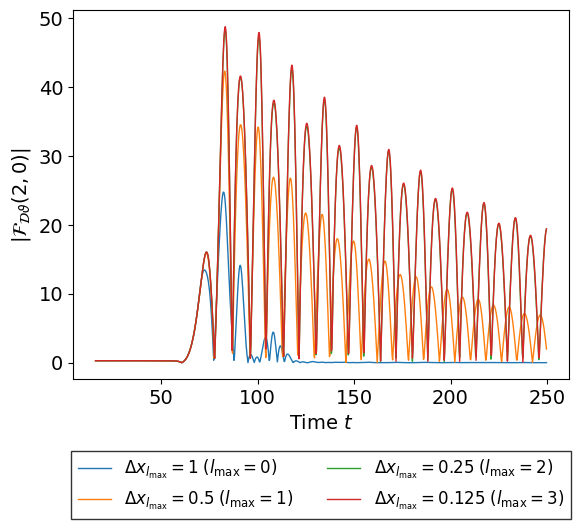}
    \caption{Absolute value of the $\{mn\} = \{2\,0\}$ Fourier mode of the massless radiation ${\cal{D}}\vartheta \cdot \hat{\bf r}$ from a $\lambda=10$ string with initial amplitude $A_0=4$, measured on a cylinder at $R=64$ for different maximum refinement levels $l_{\rm max}$ (test \textit{i)} in Table \ref{convergence_params}). The radiation is clearly not accurately captured for $\Delta x_{l_{\rm max}} \lesssim 0.25$ ($l_{\rm max}=2$) and dissipates artificially over time.}
    \label{lambda10masslessmodesnoregridquadrupole}
\end{figure}

\begin{figure}
    \centering
    \includegraphics[width=0.5\textwidth]{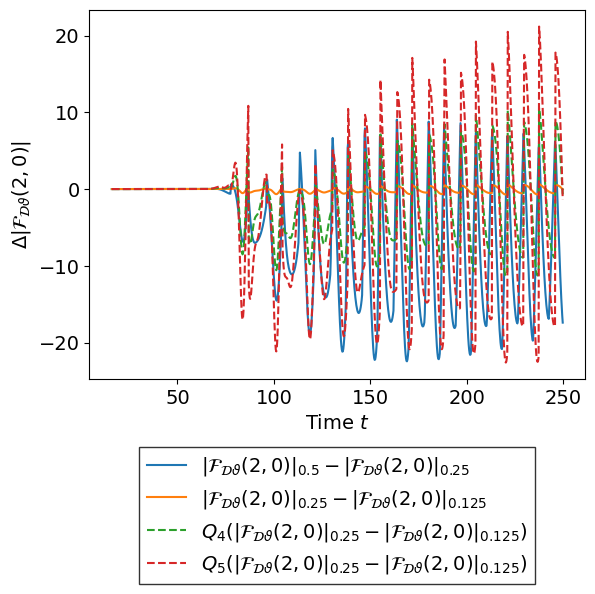}
    \caption{Convergence of the Fourier mode $|\mathcal{F}_{{\cal D}\vartheta}(2,0)|$ with grid resolutions on the finest refinement level $\Delta x_{l_{\rm max}}=0.5,\,0.25\; \rm{and}\; 0.125$ for the string configuration plotted in Figure \ref{lambda10masslessmodesnoregridquadrupole} (test \textit{i)} in Table \ref{convergence_params}). We plot the difference in the magnitude of the mode $\Delta |\mathcal{F}_{{\cal D}\vartheta}(2,0)|$ between different resolutions, with the higher resolution results also plotted rescaled according to fourth- and fifth-order convergence.}
    \label{ModeConvergence}
\end{figure}

As outlined above, an important parameter when performing string simulations is the regridding threshold $\phi_{threshold}$ for the adaptive mesh. This must be chosen so that higher levels of refinement are concentrated at the string core, in order for the string to evolve accurately and to properly resolve the outgoing radiation. This is particularly the case for higher $\lambda$, where the string radius becomes narrower and the wavelength of the lowest energy massive modes decreases. An example of the effect of insufficient regridding is given by Figure \ref{lambda10masslessmodesnoregridquadrupole}, which shows the magnitude of the massless quadrupole $\{mn\}=\{2\,0\}$ mode for different maximum refinement levels $l_{\rm max}$ emitted by a $\lambda=10$ string with initial amplitude $A_0=4$ (see Section \ref{masslessradiationanalysis} for further details about mode calculation). We observe that, for no refinement with $l_{\rm max}=0$ and $\Delta x_{l_{\rm max}}=1$, the massless radiation is not accurately resolved, artificially dissipating as the simulation progresses (along with artificial damping of the string motion, not shown). As $l_{\rm max}$ is increased, more of the massless radiation is captured, converging to a stable value at approximately $\Delta x_{l_{\rm max}}=0.25$ ($l_{\rm max}=2$) for this non-relativistic configuration. Figure \ref{ModeConvergence} shows a plot of the difference in the $\{2\,0\}$ mode between simulations of resolutions $\Delta x_{l_{\rm max}}=0.5,\, 0.25\,\; \mathrm{and} \,\;0.125$. We also plot the difference between the two finest levels multiplied by appropriate convergence factors, defined by \cite{Alcubierre}
\begin{equation}
    Q_n = \frac{(\Delta x_{\rm coarsest})^n - (\Delta x_{\rm middle})^n}{(\Delta x_{\rm middle})^n - (\Delta x_{\rm finest})^n}\,,
\end{equation}
where $n$ is the order of convergence. By comparison with the difference in the magnitude of the mode $\Delta |\mathcal{F}_{{\cal D}\vartheta}(2,0)|$ between the two coarsest levels, Figure \ref{ModeConvergence} shows that we obtain between fourth and fifth order convergence.\footnote{We note that, for this test, we used a more accurate (fourth order) extraction cylinder than described in Section \ref{diagnostictools}, which was not available when the bulk of the simulations in this paper were performed. This is not important for the results in the rest of the paper, but does affect convergence.}\textsuperscript{,}\footnote{We further note that, for convergence tests with AMR, often the resolution of the base grid $\Delta x_0$ is changed and $l_{\rm max}$ capped at a certain level i.e. as in test \textit{ii)} in Table \ref{convergence_params}. Here we take an alternative approach, fixing $\Delta x_0$ to remain the same with $l_{\rm max}$ increased for higher resolution simulations.} By fourth order Richardson extrapolation of the finest two simulations, we obtain a discretisation error estimate of $\Delta |\mathcal{F}_{{\cal D}\vartheta}(2,0)|/|\mathcal{F}_{{\cal D}\vartheta}(2,0)| \sim 0.2\%$ (in relation to the late time radiation amplitude).

\begin{figure}
    \centering
    \includegraphics[width=0.5\textwidth, trim=10 0 0 0, clip]{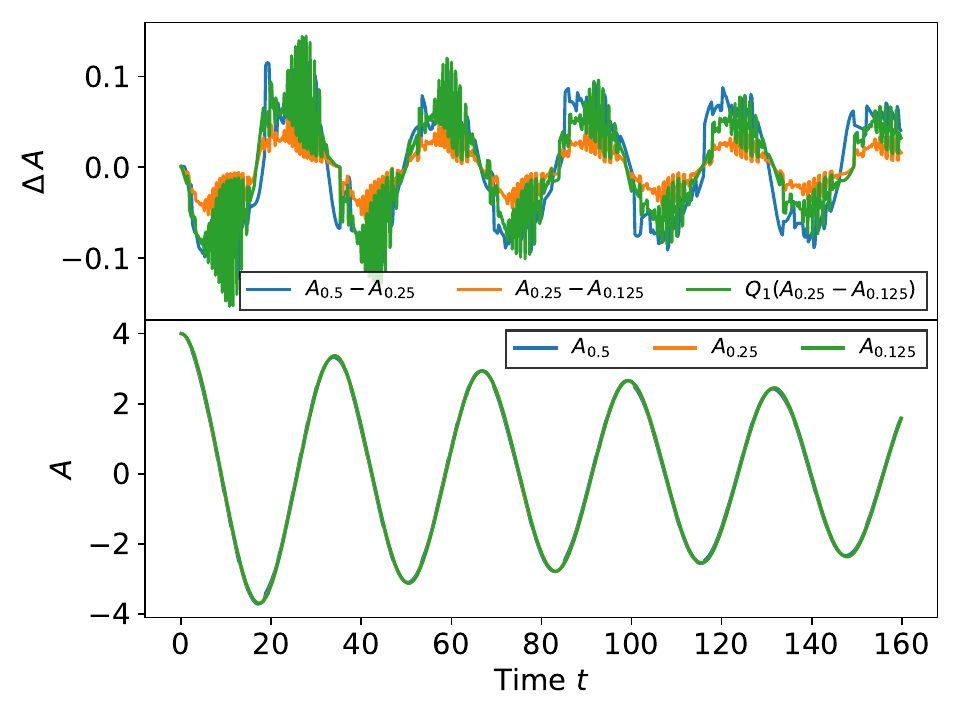}
    \caption{Convergence of the string amplitude $A$ calculated using the core finder in Section \ref{stringcoreposition}. The grid resolutions on the finest refinement level $l_{\max}=3$ are $\Delta x_{l_{\rm max}}=0.5,\,0.25\; \rm{and}\; 0.125$. The top panel shows the difference in amplitude $\Delta A$ between different resolutions, with the higher resolution results also plotted rescaled according to first-order convergence. The bottom panel shows the value of the amplitude $A$ for each simulation; the difference is indistinguishable at this scale.}
    \label{convergence2}
\end{figure}

Figure \ref{convergence2} gives the results of a convergence test using the amplitude of the string $A$. In this case, we fix $l_{\rm max}=3$ and change the resolution of the base grid, with parameters given by case \textit{ii)} in Table \ref{convergence_params}. We observe approximately first order convergence in $A$. This is what we expect, due to the dependence of the measured amplitude on $\Delta x$, as given by equation \eqref{xcore}. Using first order Richardson extrapolation, we estimate the error introduced by the discretisation to be $\Delta A / A \sim 2\,\%$ (in relation to the maximum amplitude). However, we also note that most of the error arises in the moving parts of the oscillation, rather than the maxima or minima which are used in later sections to calculate the radiation backreaction. This means the discretisation error in the backreaction calculation will be lower than this estimate, with larger sources of error coming potentially from systematic effects, such as the amount of dissipative evolution employed (see e.g. Figure \ref{fig:decayA1}).

\subsubsection{Massive Radiation Trapping and Relative Energy Loss}\label{Discussion}

\begin{figure}
    \centering
    \includegraphics[width=0.5\textwidth]{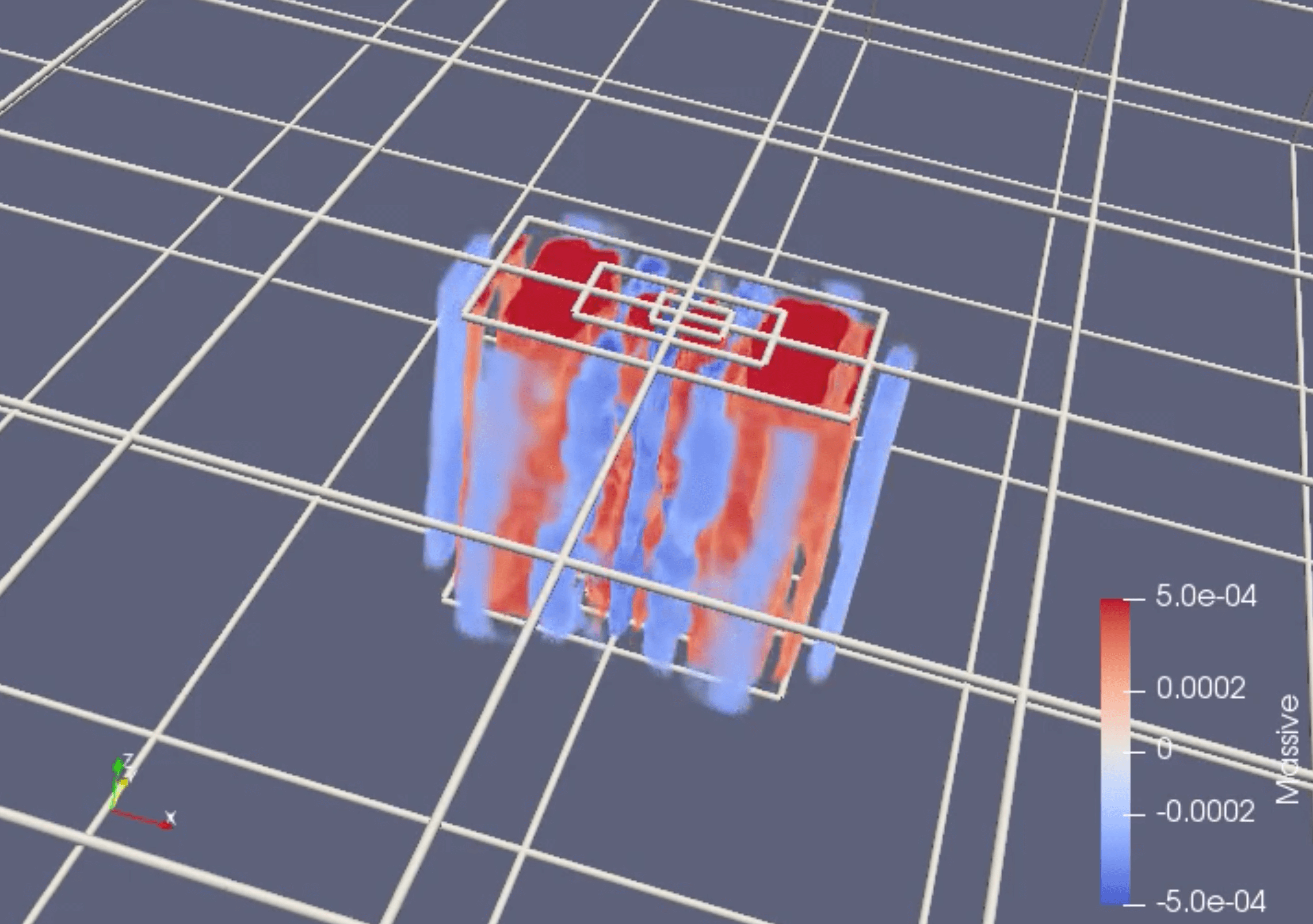}
    \caption{Volume rendering of the massive radiation $\Pi_\phi$ from a $\lambda = 3$ string, with AMR blocks outlined in white. Much of the radiation signal is trapped at the AMR boundary, and this short wavelength signal grows in amplitude over time.}
    \label{TrappedRadiation}
\end{figure}

Through running a large number of simulations of global strings, we observe that massive radiation is very sensitive to the grid resolution and can be significantly affected by the adaptive remeshing.  This manifests at higher $\lambda$ by high frequency massive modes becoming `trapped' on finer grids near the string, a process which can lead to further radiation growth due to stimulated emission or resonance. This is due to the creation of internal reflections from the AMR boundaries where the grid refinement steps down, causing a resonant effect which accumulates over time, as shown in Figure \ref{TrappedRadiation}. Further to this, we observe that the remeshing of the grid itself can introduce artificial massive radiation, due to the inevitably slightly inaccurate interpolation between finer and coarser refinement levels. These standing wave instabilities are a familiar shortcoming of AMR for which remedies include the introduction of artificial dissipation by using Kreiss-Oliger damping. 

\begin{figure}
    \centering
    \includegraphics[width=0.5\textwidth]{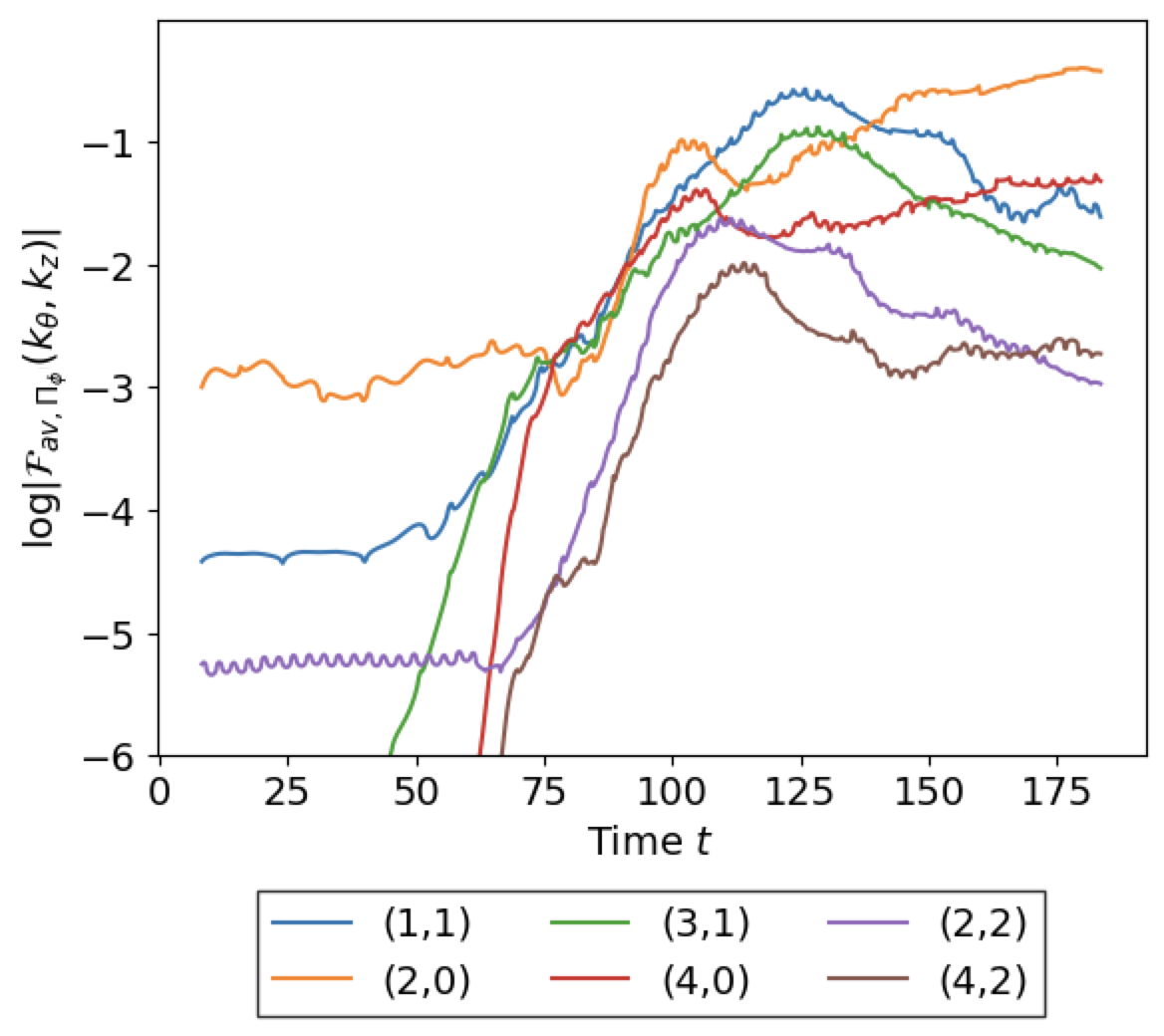}
    \caption{Dominant 2D Fourier modes of the massive radiation $\Pi_\phi$ from a $\lambda=1$ string with initial amplitude $A_0=4$, measured on a cylinder at $R=64$ and time averaged over approximate half-period $\Delta t/2 = 66/4$.}
    \label{temp2}
\end{figure}

In practice, the energy loss from massive radiation can be ignored for the quasilinear string configurations studied in this paper. For example, at low amplitude $A_0=1$ ($\varepsilon \approx 0.2$) and $\lambda=1$, we measure the amplitude of the dominant massive mode to be $\sim 10^5\times$ smaller than the dominant quadrupole massless radiation signal, so it has a negligible effect on string motion and should not influence our massless results. We also note that even for quasinonlinear intermediate amplitude regimes with $A_0=4$ ($\varepsilon \approx 0.7$), the magnitude of the massive signal is $\sim 10^3\times$ smaller than the massless emission, so the energy losses remain very small indeed. This is demonstrated by comparing the massive modes in Figure \ref{temp2} to the massless modes from the same configuration in the middle panel of Figure \ref{masslessspatialamp1regridding0.25cumulative}. (Further information about how the modes are calculated is given in Section \ref{masslessradiationanalysis}.) In both cases, strings therefore radiate preferentially into massless channels and alternative massive channels are strongly suppressed. Further investigation of these radiative effects into the nonlinear regime will be addressed in future work \cite{Drew2020}.

\section{Massless (Axion) Radiation}\label{Massless}

\subsection{Analytic Radiation Expectations}

\subsubsection{Separable Radiation Eigenmodes}\label{separableradiation}

\begin{figure*}
    \centering
    \includegraphics[width=0.75\textwidth]{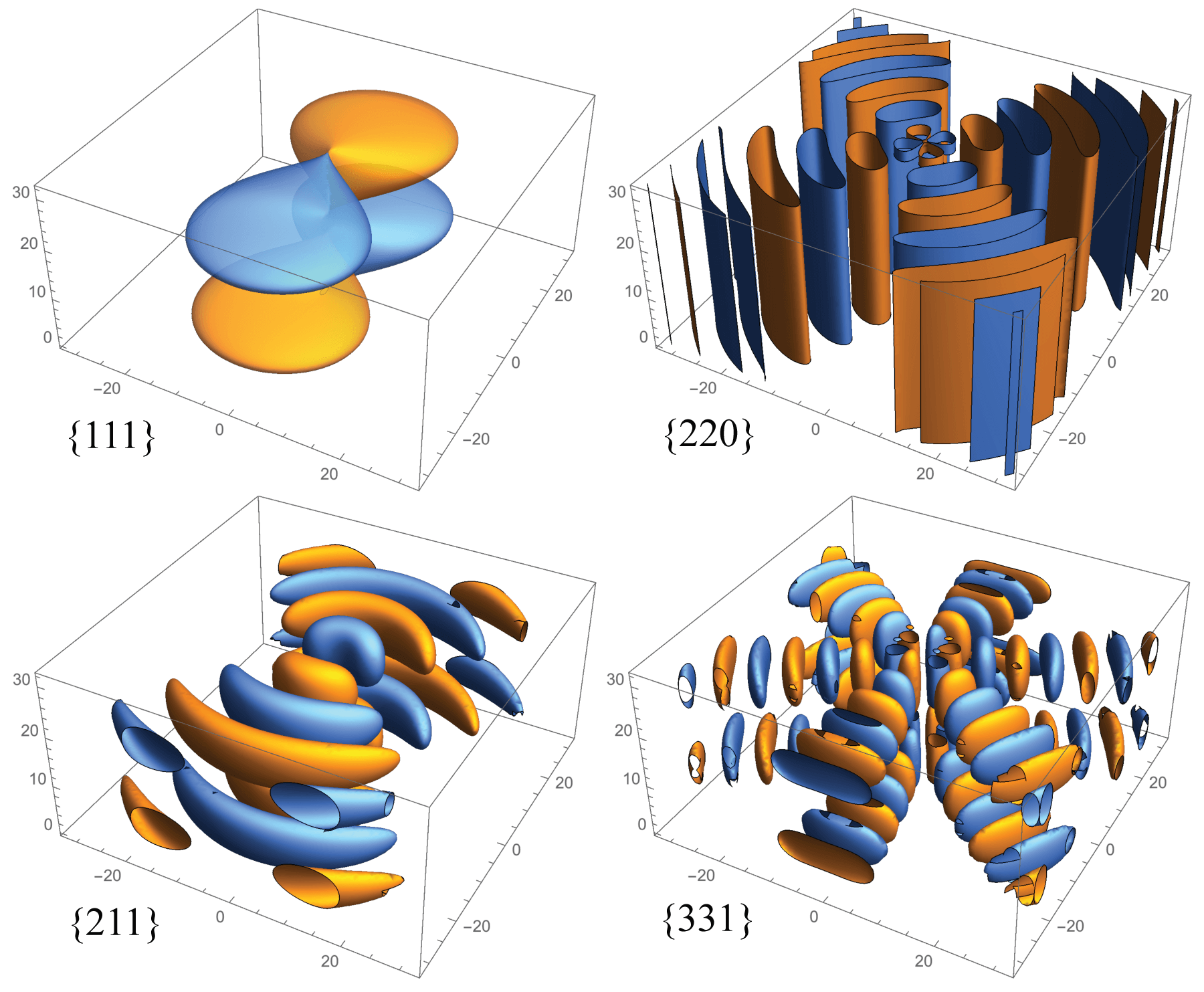}
    \caption{Key radiation eigenmodes for a periodic oscillating string as labelled by their eigenvalues $\{t,\,\theta,\, z\}\rightarrow \{pmn\}$ with the key radial eigenvalue $\kappa_{pn}$ given by the dispersion relation (\ref{dispersionrelation}). The oscillating string self-field creates a non-propagating (evanescent) wave with $\{pmn\}=\{1\,1\,1\}$ (top left), while the dominant massless radiation mode is the quadrupole $\{2\,2\,0\}$ (shown top right).  The next most important massless radiation mode is the third harmonic $\{3\,3\,1\}$ (bottom right) but there appears to be some contribution from a second harmonic dipole $\{2\,1\,1\}$ (bottom left). For massive scalar radiation, higher harmonic dipole modes $\{p\,1\,1\}$ can provide the leading contribution, but they compete with quadrupole $\{p\,2\,2\}$ and other higher modes. }
    \label{fig:fig_radiationmodes}
\end{figure*}

To guide our analysis of the massless radiation emitted by an oscillating string, we shall assume as in Section \ref{masslessmassive} that we are far away from the source with the field very close to the vacuum state $\phi=1$.  Removing massive excitations in this way, we obtain the massless wave equation (\ref{masslessradiation}) which in cylindrical coordinates  $\vartheta(t,r,\theta,z)$ becomes 
\begin{equation}\label{masslesswaveeqn}
\frac{\partial ^2 \vartheta}{\partial t^2} - \frac{\partial ^2 \vartheta}{\partial r^2}- \frac{1}{r}\frac{\partial  \vartheta}{\partial r} - \frac{1}{r^2}\frac{\partial ^2 \vartheta}{\partial \theta^2} - \frac{\partial ^2 \vartheta}{\partial z^2}  = 0\,.
\end{equation}
Taking a periodic oscillating string along the $z$-axis (with $0\le z< L$), the massless radiative modes emitted will become outgoing solutions of (\ref{masslesswaveeqn}) at large distances. We can solve the cylindrical wave equation via separation of variables with the \textit{ansatz}
\begin{eqnarray}
&&\vartheta(t,r,\varphi ,z) = T(t)\,R(r)\, \Theta (\theta )\, Z(z) ~~~~ \Rightarrow  ~~~~
\label{separation}\\ \nonumber\\
&&\frac{T''(t)}{T(t)}-  \frac{R''(r)+ R'(r)/r}{R(r)}-\frac{1}{r^2 }\frac{\Theta ''(\theta )}{\Theta (\theta )} - \frac{Z''(z)}{Z(z)}=0\,,\nonumber
\end{eqnarray}
where each component is solved in turn by introducing appropriate separation constants. The time dependence $T(t)$ is given by the period of the string oscillations (or their $p$th harmonics) with angular frequency 
\begin{equation}
\omega_p = \frac{2\pi}{L}\frac{p}{\alpha} \;\equiv\; \Omega_z \, \frac{p}{\alpha}\,,
\end{equation}
where $p$ is a positive integer and the parameter $\alpha\ge 1$ is the fractional increase in the path length of the string as it is traverses  from $z=0$ to $z=L$.  The string oscillates with a period $T\equiv \alpha L \gtrsim L$ determined by its actual invariant length, where for small relative amplitude ($\varepsilon \rightarrow 0$) we have $\alpha\approx1$ (see the next subsection).  The time-dependence of the separable solutions becomes
\begin{equation}\label{timeoutgoing}
 T_p(t) ~\propto~ \, e ^ {- i\,\Omega_z\,p \,t/\alpha} \,,
\end{equation}
where we take the convention of a negative sign for the outgoing mode.  The fixed periodicity along the $z$-axis (length $L$) yields a further separation constant from ${Z''/Z}  =  -k_z^2 = - \Omega_z^2 n^2$, giving the eigenmode
\begin{equation}
Z_n(z) ~\propto~ e ^ {i \Omega_z nz}\,,
\end{equation}
where $n$ is an integer and the wavenumber in the $z$-direction is $k_z = \Omega_z n$ with $\Omega_z = 2\pi/L$.   

Substituting these eigenmodes into equation (\ref{separation}), we obtain the $\theta$-dependence
\begin{equation}
    \frac{\Theta ''(\theta )}{\Theta (\theta )}=- r^2\left(\frac{R''(r)+ R'(r)/r}{R(r)} +\omega_p^2 - k_z^2 \right)\,,
\end{equation}
where the right hand side is independent of $\theta$ and can be set to a constant. The azimuthal periodicity of $\theta$ (period $2\pi$) gives $\Theta(\theta) ''/\Theta(\theta) = -m^2$, with the $\theta$-dependence
\begin{equation}
    \Theta_m (\theta ) ~\propto~ e ^ {i m \theta}\,,
\end{equation}
where $m$ is an integer. A final rearrangement gives 
\begin{equation}
R''(r) + \frac{R'(r) }{r} + R(r) \left(\omega_p^2-k_z^2 -\frac{m^2}{r^2}\right)=0\,,
\end{equation}
where we identify the radial wavenumber $k_r^2 = \omega_p^2 - k_z^2$. Factoring out the dependence on $\Omega_z$ from this dispersion relation and setting $k_r=\Omega_z\kappa_{pn}$, we obtain the important expression for the radial wavenumber for each harmonic
\begin{equation}\label{dispersionrelation}
\kappa_{pn} = \sqrt{\left({p}/{\alpha}\right)^2 - n^2}\,.
\end{equation}
We note that the radial wavenumber depends on $p$ and $n$ only, and that the angular dependence $m$ has decoupled. This finally leaves the radial Bessel's equation,
\begin{equation}\label{radialbesseleqn}
R''(r) + \frac{R'(r) }{r} + R(r) \left( \Omega_z^2\kappa_{pn}^2 -\frac{m^2}{r^2}\right)=0\,,
\end{equation}
which has solutions which are arbitrary linear combinations of Bessel functions of the first kind $J_m(k_r r)$ and second kind $Y_m\left(k_r r\right)$.  However, when we impose the Sommerfeld radiation condition, 
\begin{equation}
r^{1/2} \left ( \frac{\partial}{\partial r} - i k_r \right) \vartheta \;\rightarrow 0 ~~\hbox{as}~~r\rightarrow \infty \,,
\end{equation}
the solution is constrained to be a Hankel function of the first kind, with 
 \begin{equation}\label{radialsoln}
R_{pmn}(r) ~\propto~ H^{(1)}_{m}(\Omega_z\kappa_{pn}) = J_m(\Omega_z\kappa_{pn}r)+i Y_m(\Omega_z\kappa_{pn}  r)\,.
\end{equation}
By comparing with the time-dependence (\ref{timeoutgoing}), the outgoing mode from the asymptotic behaviour at large radial distances $\kappa_{n\ell} r\gg 1$ is given by
\begin{equation}
H^{(1)}_{m}(k_r r) \approx \sqrt{{2}/{\pi k_r r}} \,\exp\left[i \left(k_r r -{\pi  m}/{2}-{\pi }/{4}\right)\right]\,.
\end{equation}
We note that there is an apparent divergence at $r=0$ for $Y_m\left(k_r r\right)$, but that this can be modified and cut off by the near-field dynamics and structure of the global string. Here, we are only seeking the matching asymptotic solution for the far-field with $k_r r\gg1$. Finally, combining these results, we find the general outgoing radiation solution from a sum over the separable modes:
\begin{eqnarray}\label{generalsolution}
\vartheta (t,r,\theta ,z) &=&  \Re \sum_{p\,m\,n} A_{pnm} \,   \;e^{i m \theta}  \\
&&~\times ~ \, e^{- i\,\Omega_z \,\left[(p/\alpha)\,t - nz \right]} \; H^{(1)}_m\left(\Omega_z\, \kappa_{pn} \,r\right)\,.\nonumber
\end{eqnarray}
with  amplitude $A_{pmn} $ for the specific $\{t,\theta,z\}$ eigenmode labelled by the integers $\{pmn\}$ and with the key radial eigenvalue $\kappa_{pn}$  given by (\ref{dispersionrelation}).

We can make several observations about the radiation solution (\ref{generalsolution}) using the associated dispersion relation (\ref{dispersionrelation}). First, to aid with interpretation of the discussion, Figure \ref{fig:fig_radiationmodes} shows some of the most significant eigenmodes for string radiation $\{pnm\}$ = $\{1\,1\,1\}$, $\{2\,0\,0\}$, $\{2\,1\,1\}$ and $\{3\,3\,1\}$.  As we shall see in the next section, the sinusoidal solution (\ref{sineapprox}) has a long-range self-field which oscillates backwards and forwards with the string which can be associated with the eigenmode $\{pmn\} = \{1\,1\,1\}$.  This is an apparent dipole, but it is not  true radiation and it will not propagate in the outward direction because the radial wavenumber is imaginary, i.e.\ $\kappa_{11}^2 <0$.  This self-field contribution is therefore an evanescent wave with no net flux through our diagnostic radiation cylinder when averaged over a full oscillation period.  The radiation mode predicted to be dominant is the quadrupole $\{2\,2\,0\}$, which propagates radially at the speed of light with $\omega_2 = k_r = \Omega_z \kappa_{20}$, where $\kappa_{20} = 2/\alpha \approx 2$ at small amplitude.   In principle, the second harmonic $\{2\,1\,1\}$ can also propagate, but in practice we find that the third $\{3\,3\,1\}$ and fourth $\{4\,4\,0\}$ harmonics make the next most important contributions. The mode amplitudes are determined by the dynamics and symmetries of the near-field physics of the specific configuration of the oscillating string source. Finally, the dispersion relation (\ref{dispersionrelation}) also reveals that not all modes propagate at the speed of light in the radial direction (Huygen's principle does not work in 2D).  If $n\ne 0$, then there is a wavevector component in the $z$-direction and the radial speed of propagation is $v_r = \partial \omega /\partial k_r = \kappa_{pn}/\omega_p < 1$. For example, for $\{2\,1\,1\}$, we have $v_r = 0.87$, while for $\{3\,3\,1\}$ we have $v_r = 0.94$.

\subsubsection{Separation of Propagating Modes from Self-Field}
\label{radvsself}
\begin{figure}
    \centering
    \hspace*{-0.4cm}
    \includegraphics[width=0.98\linewidth, trim=0 80 0 0, clip]{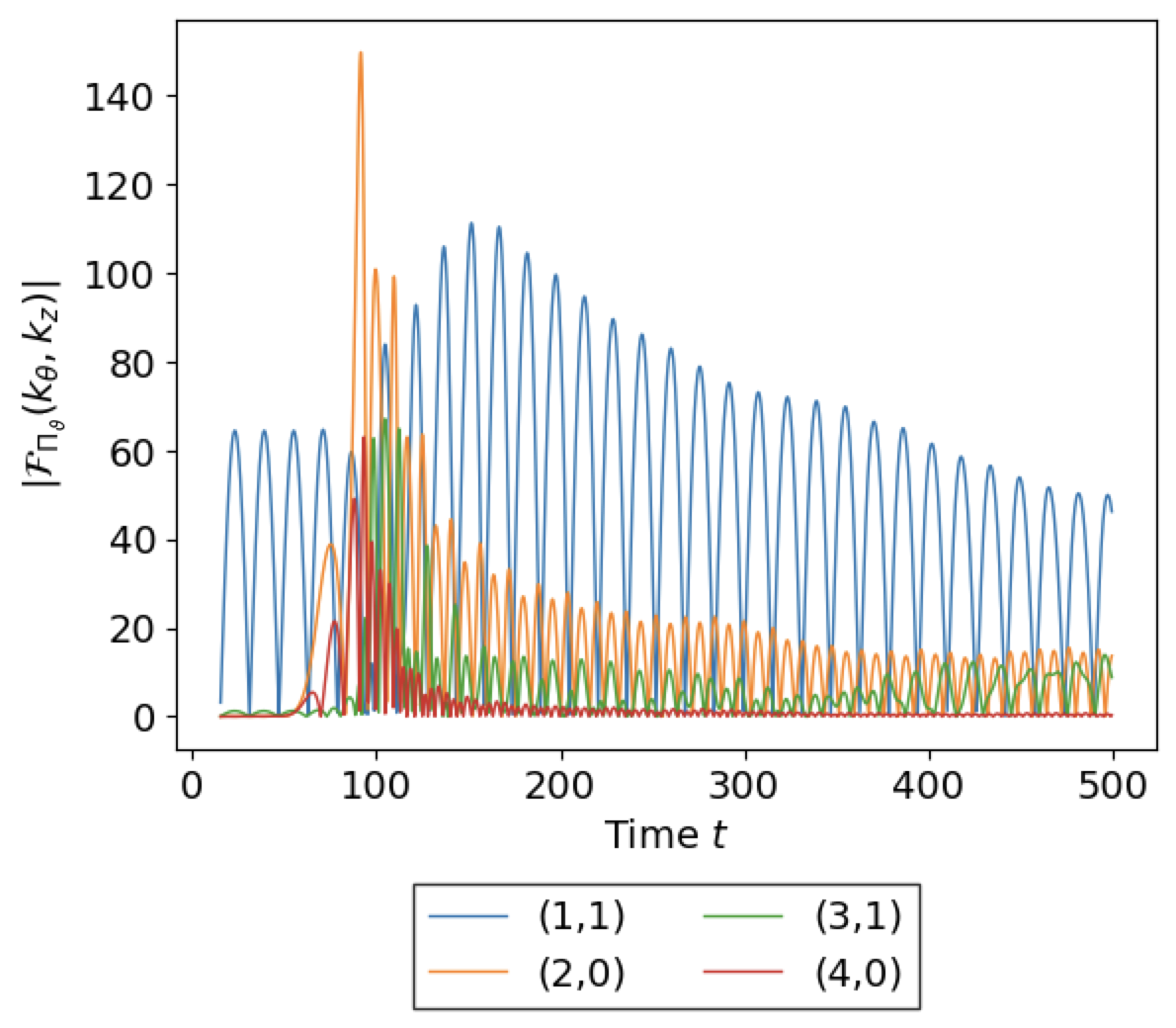}
   \includegraphics[width=\linewidth]{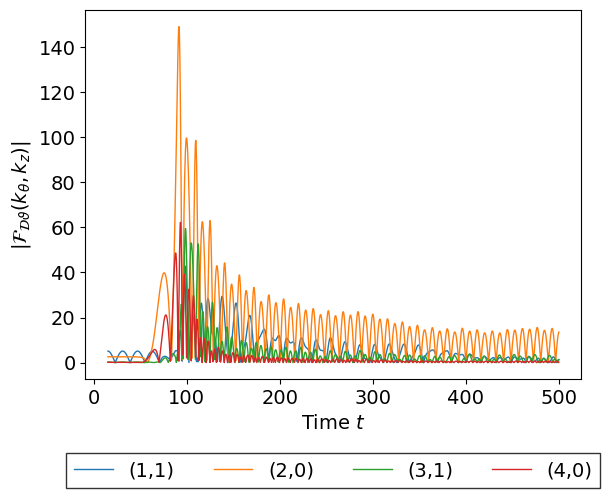}
    \caption{Absolute value of the $\{mn\} = \{1\,1\}, \{2\,0\}, \{3\,1\},$ and $\{4\,0\}$ Fourier modes of the massless radiation $\Pi_\vartheta$ (top) and ${\cal{D}}\vartheta \cdot \hat{\bf r}$ (bottom) from a $\lambda=1$ string with initial amplitude $A_0=12$, measured on a cylinder at $R=64$.}
    \label{selffieldreduction} 
\end{figure}

As discussed in Section \ref{masslessmassive}, the study of string radiation is plagued by long-range self-fields that prove difficult to numerically separate.  Given our fixed cylinder at a finite distance $R$ from the string, there is a prosaic explanation for this contamination due to the oscillating self-fields being offset from their central position (rather than the evanascent waves described above). At small amplitude ($\varepsilon \ll 1$), the sinusoidal string solution (\ref{sineapprox}) with the string field \textit{ansatz} (\ref{phi}) yields an approximate massless self-field $\vartheta_{\rm sf}(t, \bf{x})$ of the following form:
\begin{eqnarray}
\label{selffieldapprox}
\vartheta_{\rm sf}(t,{\bf x}) \approx \tan^{-1} \left ( y/ X(t,{\bf x}) \right)\,,\nonumber\\
X(t,{\bf x}) = x - A \cos\Omega_z t \, \sin\Omega_z z\,, 
\end{eqnarray}
which should be valid in the region $A\ll r \lesssim {\cal O}(\hbox{few}\times L)$.   Taking the time derivative of the oscillating field $\vartheta_{\rm sf}$, we find on a cylinder at a distance $r=R$ that to leading order 
\begin{eqnarray}
\label{selffielddipole}
\frac{\partial \vartheta_{\rm sf}}{\partial t}\left (t, r, \theta, z\right ) \, &\approx&\,  \frac{A \Omega_z}{R} \sin\theta\,  \cos\Omega_z t \, \sin\Omega_z z\,, 
\end{eqnarray}
so the non-propagating self-field at a fixed radius looks like a dipole field. (This corresponds to the mode $\{1\,1\,1\}$ derived in the previous section, also see Figure \ref{fig:fig_radiationmodes}).  The radial derivative of $\partial \vartheta_{\rm sf}/\partial r$ yields the same dipole space and time dependence as (\ref{selffielddipole}), except that the pre-factor becomes $A/R^2$ so the amplitude falls off more steeply with distance than the time derivative $\dot\vartheta_{\rm sf}$.  This means that the spatial radiation diagnostic $\cal{D}\vartheta$ in (\ref{masslessdiagnostic}) is more effective for removing self-field contamination than the massless field momentum ${\Pi}_\vartheta $. An example of this is shown in Figure \ref{selffieldreduction}, which shows a plot of the absolute value of the $\{mn\} = \{1\,1\}, \{2\,0\}, \{3\,1\},$ and $\{4\,0\}$ Fourier modes of the two massless radiation diagnostics from a $\lambda=1$ string with initial amplitude $A_0=12$. We clearly observe that ${\Pi}_\vartheta$ is significantly more contaminated by the $\{1\,1\}$ self-field signal, and that $\cal{D}\vartheta$ almost entirely removes this signal without affecting the other modes. This diagnostic is also therefore a very useful quantity for visualisation, producing much cleaner massless radiation signal.

\subsubsection{Dual Radiation Calculations and String Backreaction}\label{dualradiation}

There is a well-known duality between the massless Goldstone boson $\vartheta$ in the $U(1)$ model (\ref{GoldstoneLagrangian}) and a two-index antisymmetric tensor $B_{\mu\nu}$ through the relation \cite{Vilenkin:1986ku,Davis:1988rw}
\begin{equation}
\label{duality}
\phi^2\partial_{\mu}\vartheta = \frac{1}{2}f_a\epsilon_{\mu\nu\lambda\rho}
\partial
^{\nu}B^{\lambda\rho}\,.
\end{equation}
After integrating radially over the massive degrees of freedom, this alternative description yields the Kalb-Ramond action which consists of the familiar Nambu action for a local string coupled to the antisymmetric tensor $B_{\mu\nu}$ \cite{Kalb:1974yc}.   This is closely analogous to the coupling of a local string to the gravitational field and allows direct calculation of the resulting propagating radiation fields.  Specific linearised solutions have been obtained for axion radiation from both closed loops \cite{Vilenkin:1986ku} and the long string solutions being considered here, see \cite{Sakellariadou:1991sd} and \cite{Battye1993}. We will not repeat these calculations, only recounting the key results from ref.~\cite{Battye1993}. 

For small amplitude $\varepsilon \ll 1$, our sinusoidal long string initial condition (\ref{sineapprox}) in a box of length $L$ approximates an analytic solution of the Nambu-Goto equations of motion with time period $T\approx L$, which we can parametrise in terms of left-moving ($u=\sigma+t$) and right-moving ($v=\sigma-t$) coordinates along the string.  Taking the relative amplitude $\varepsilon \equiv 2\pi A/T = \Omega A$, where $\Omega = 2\pi/T$, this periodic solution takes the form 
\begin{equation}
\label{sineanalytic}
{\bf X} = \bigg{(}\frac{\varepsilon}{2\Omega}\big{[}\cos\Omega u +
\cos\Omega v
\big{]},0,\frac{1}{2\Omega}\big{[}{\mathsf E}(\Omega u, \varepsilon) +
{\mathsf E}(\Omega v, \varepsilon)
\big{]}\bigg{)}\,.
\end{equation}
Here, ${\mathsf E}(\phi, \varepsilon)$ is the incomplete elliptic integral of the
second kind
\begin{align}
\label{}
{\mathsf E}(\phi, \varepsilon)&=\int_0^{\phi}d\theta\sqrt{1-\varepsilon^2\sin ^2\theta} \qquad (\varepsilon \le 1) \nonumber \\
&\approx~ \frac{2}{\pi}\left (\phi\, {\mathsf E}(\varepsilon) + \frac{\varepsilon^2}{8} {\mathsf K}(\varepsilon) \sin(2 \phi) + ...\right) 
\end{align}
where ${\mathsf E}(\varepsilon) \equiv {\mathsf E} (\pi/2, \varepsilon) = \pi/2(1 - \varepsilon^2/4 - 3 \varepsilon^4/64 + ...) $ and ${\mathsf K}(\varepsilon) = \pi/2(1 + \varepsilon^2/4 + 9 \varepsilon^4/64 + ...)$ are the complete elliptic integrals of the second and first kind respectively. Due to the non-zero amplitude $\varepsilon$, the spatial coordinate $\sigma$ measuring the invariant length along the string is no longer directly proportional to the $z$-coordinate of the numerical grid, giving $
z \approx  2 {\mathsf E}(\varepsilon) \sigma / \pi +  \hbox{periodic terms} \approx \sigma (1 - \varepsilon^2/4 - ...)\,
$.  Imposing spatial periodicity $L$ in the $z$-direction, the energy of the string in the same interval (setting $\sigma = T$) becomes 
\begin{eqnarray}
E(\varepsilon) &\equiv& \mu T(\varepsilon) =\frac{\pi} { 2 {\mathsf E}(\varepsilon)} \mu L \nonumber \\ 
\label{sineenergy}
&\approx&  \mu L \left ( 1+ \textstyle{\frac{1}{4}}\varepsilon^2 + \textstyle{\frac{7}{64}}\varepsilon^4 + ...\right )\,.
\end{eqnarray}
As $\varepsilon$ increases, the true time periodicity $T$ differs from the spatial periodicity $L$ (with $T = \alpha L > L$). The increase in periodicity is illustrated by the oscillating string in Figure \ref{amplitudedecay} with $A_0=4 $ $(\varepsilon=0.68)$, which shows an initial periodicity about 11\% longer than $L=32$ (at large $\lambda$), though this is lower than the expected 15\% due to long-range forces and radiative backreaction accelerating the string (and modifying $\varepsilon$).   Relativistic effects become important as $\varepsilon \rightarrow 1$ ($A_0=8$) with $T \rightarrow \pi L/2$ and, in this limit, two points along the string approach a cusp ($v\rightarrow 1$) twice each period. In this paper, most quantitative tests will be undertaken at smaller $\varepsilon$ where we can neglect these corrections. 

Massless radiation calculations using the antisymmetric tensor formalism (\ref{duality}) have been undertaken for periodic solutions like (\ref{sineanalytic}), with the radiation power $P$ expressed as a sum over the $n$ harmonics $P_n$, generally yielding combinations of Bessel functions.   A particularly interesting case is the periodic helix for which a full nonlinear analysis can be performed \cite{Sakellariadou:1991sd}, showing that only harmonics with $m+n$ \textit{even} radiate, with large $n$ harmonics exponentially suppressed $P_n\propto e^{-\zeta n}$, where $\zeta$ is larger at small relative amplitude $\varepsilon$ \cite{Battye1993}.  Symmetry prevents the helix from radiating in the generic $n=2$ quadrupole mode (the lowest harmonic is $n=3$). We focus here instead on analytic calculations for the sinusoidal solution (\ref{sineanalytic}).  In this case, a linearized calculation of the leading-order radiation from the second harmonic yields a power per unit length of \cite{Battye1993}
\begin{equation}
\label{sineradiation}
\frac{dP}{dz} = \frac{\pi^3\eta^2}{16L}\varepsilon^4 \,.
\end{equation}
The same calculation applied to more realistic configurations with a superposition of sinusoidal modes (even for a kink solution i.e. a solution with a discontinuous tangent vector) also has the leading $P \propto \varepsilon^4$ dependence, summed over the contributing modes.  

\begin{figure}[!t]
    \centering     
    \includegraphics[width=0.5\textwidth]{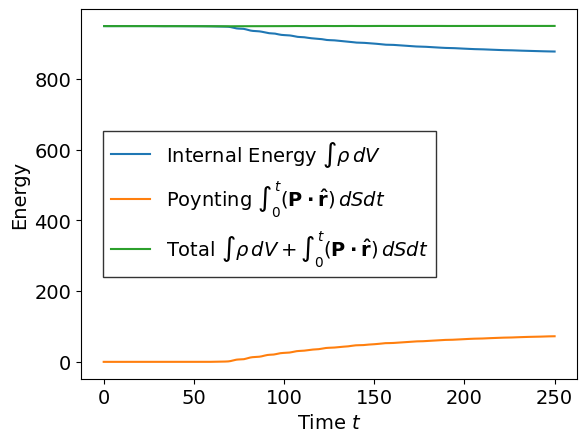}
    \caption{Comparison of the total energy within the volume enclosed by the diagnostic cylinder $S$ (blue line) with the outgoing massless radiation energy determined by the time integral of ${\cal D}\vartheta \cdot \hat{\bf r}$ (orange line). The dominant energy loss mechanism is massless radiation as indicated by conservation of the sum (green line). Conserved to $\pm 0.1\%$ after $t=250$ (1000 timesteps).}
    \label{Poyntingvectorintegrated}
\end{figure}

This generic quadrupole radiation rate can be developed into a simple analytic backreaction model describing the effect of radiation energy losses on string motion \cite{Battye1993}.  At small amplitude, we can see from (\ref{sineenergy}) that the oscillation energy to leading order is given by the square of the amplitude $\varepsilon$, where $E = \mu L (1 + \textstyle{\frac{1}{4}}\varepsilon^2)$.  Equating the rate of energy loss with the radiation power (\ref{sineradiation}) yields the time-derivative of the energy per unit length, 
\begin{equation}
\frac{1}{L}\frac{dE}{dt} = \frac{\mu}{4} \frac{d(\varepsilon^2)}{dt} =  - {\pi^3\eta^2(\varepsilon^2)^2 \over 16L}\,.
\end{equation}
This can be easily integrated to obtain the solution for the relative amplitude, 
\begin{eqnarray}
\label{inversesquare}
{1\over \varepsilon^2} - {1\over \varepsilon_0^2} = \frac{\beta \,t}{\bar \mu L}  ~~~~ \Rightarrow ~~~\varepsilon = \varepsilon_0\left ( 1 + \frac{ \beta \varepsilon_0^2 t }{\bar \mu L}\right )^{-1/2} \,,
\end{eqnarray}
where $\beta = \pi^3/4$ and $\bar \mu = \mu /\eta^2 \approx 2\pi \ln (\sqrt{\lambda} \eta R)$, where the cutoff $R$ is related to the curvature radius of the string (see earlier discussion).   Note that (\ref{inversesquare}) is a direct analytic prediction for the damping rate of a global string as a function of scale, which we will test numerically.  In evaluating whether an oscillating string conforms with this model, it is easiest to use the first expression in (\ref{inversesquare}), seeking a simple linear relation between time $t$ and the inverse square of the relative amplitude $\varepsilon ^{-2}$.

The sinusoidal solution (\ref{sineanalytic}) assumes left- and right-moving modes of equal magnitude. It has been argued, when these are unequal, that exponential decay may be more typical of radiation damping processes \cite{Battye1995}.  For our purposes, it is useful to have a second alternative model with which to compare the interpretation of results.   In principle, cross-coupled modes can cause amplitude decay like that of a damped simple harmonic oscillator, so by analogy with the power law decay in (\ref{inversesquare}) we consider the form:
\begin{equation}
\label{exponential}
\varepsilon = \varepsilon_0 \exp \left (- \frac{ \beta  t }{2 \bar \mu L}\right ) \,.
\end{equation}
Again we will test the model by seeking a linear relation, here between the time $t$ and $\ln \varepsilon $. We can also introduce an amplitude dependence (for example, see \cite{Battye1993}), so the damping rate becomes $\beta \varepsilon^2/2\bar\mu L$, where $\varepsilon^2 = |\varepsilon_L^2 - \varepsilon_R^2|$ represents the difference in amplitude between left- and right-moving modes. 

\begin{figure}[!]
    \centering
    \includegraphics[width=0.45\textwidth]{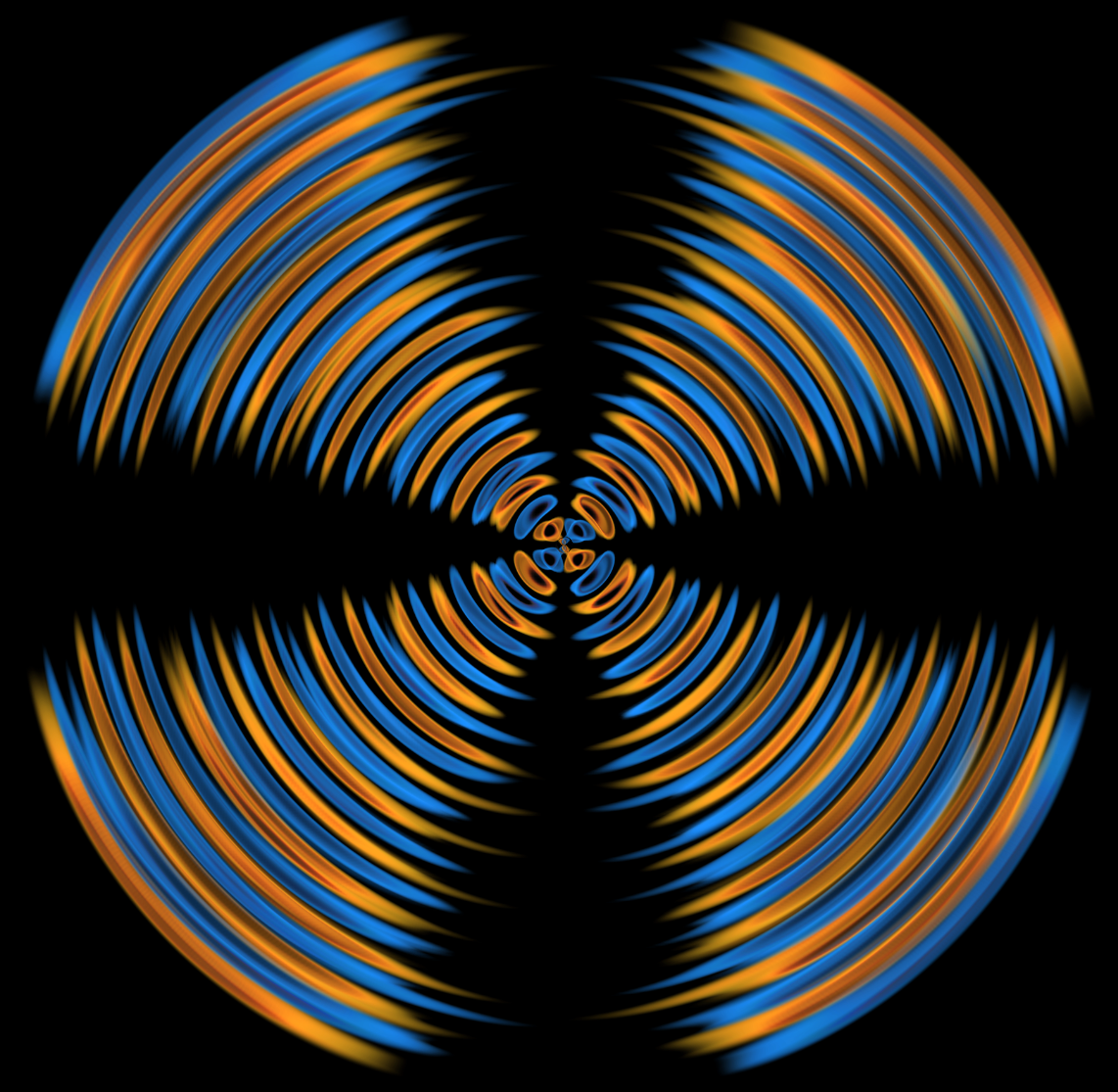}
   \includegraphics[width=0.45\textwidth]{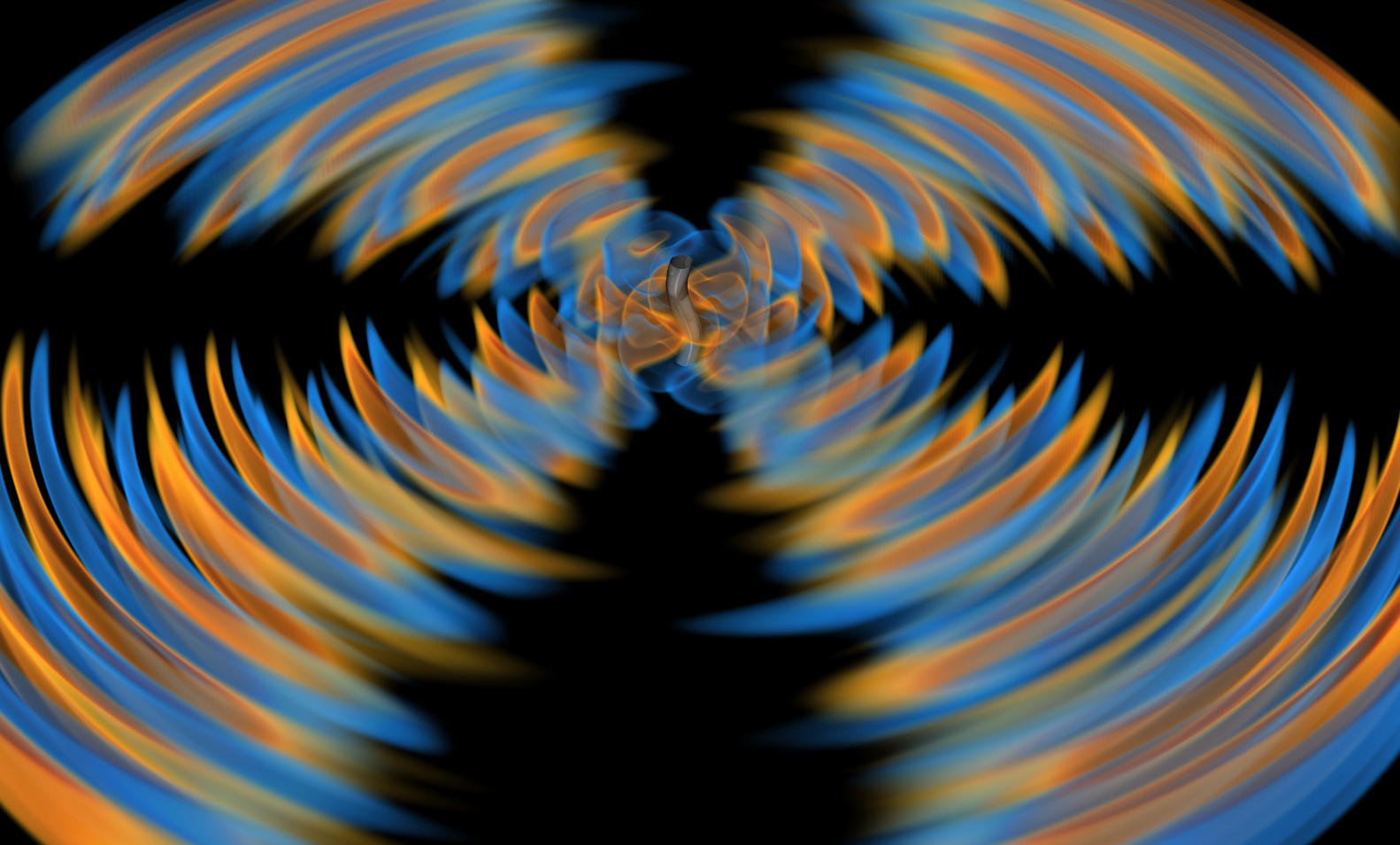}
    \caption{Volume rendering in 3D space $(x,y,z)$ of the massless radiation ${\cal{D}}\vartheta \cdot \hat {\bf r}$ from a $\lambda=1$ string with initial amplitude $A_0=4$. The radiation is emitted from a string at the centre of the grid, with the quadrupole mode $\{mn\} = \{2\,0\}$ clearly dominant.}
    \label{ParaviewMasslessScreenshot}
\end{figure}

\subsection{Massless Radiation Analysis}\label{masslessradiationanalysis}

In this section, we present a quantitative analysis of the massless radiation from oscillating string configurations with small amplitude $A_0=1$ and larger amplitudes $A_0=4$ and $A_0=8$. Simulations are set up as described in Section \ref{SimulationSetup}, and we investigate the cases $\lambda=1$ and $\lambda=10$. We analyse propagating massless radiation using the spatial diagnostic ${\cal D}\vartheta \cdot \hat{\bf r}$ defined by (\ref{masslessdiagnostic}). 

As an initial accuracy check and to establish energy conservation, we integrate the energy within the cylindrical volume enclosed by $R=64$ and the net massless radiation energy propagating across the cylinder using the interior density $\rho$ and the time integral of the massless component of the radial radiation Poynting vector $(\Pi_\vartheta{\cal D}\vartheta) \cdot \hat{\bf r}$. An example is shown in Figure \ref{Poyntingvectorintegrated} for a $\lambda=1$ string with initial amplitude $A_0=4$. This confirms accurate energy conservation for the simulation and the dominance of massless radiation losses at small amplitude.

To aid with physical understanding prior to the upcoming detailed discussion, we first present Figure~\ref{ParaviewMasslessScreenshot}, which shows a 3D spatial visualisation of the massless radiation for $\lambda=1$ with an intermediate amplitude $A_0=4$. This late-time snapshot clearly shows the dominant quadrupole structure as predicted analytically by the solution to the massless wave equation (\ref{generalsolution}). Detailed quantitative analysis of the different configurations is performed in subsequent sections by extracting and Fourier decomposing the massless radiation field over time on a cylinder at fixed radius $R=64$.

\begin{figure}
    \centering
    \includegraphics[width=0.5\textwidth]{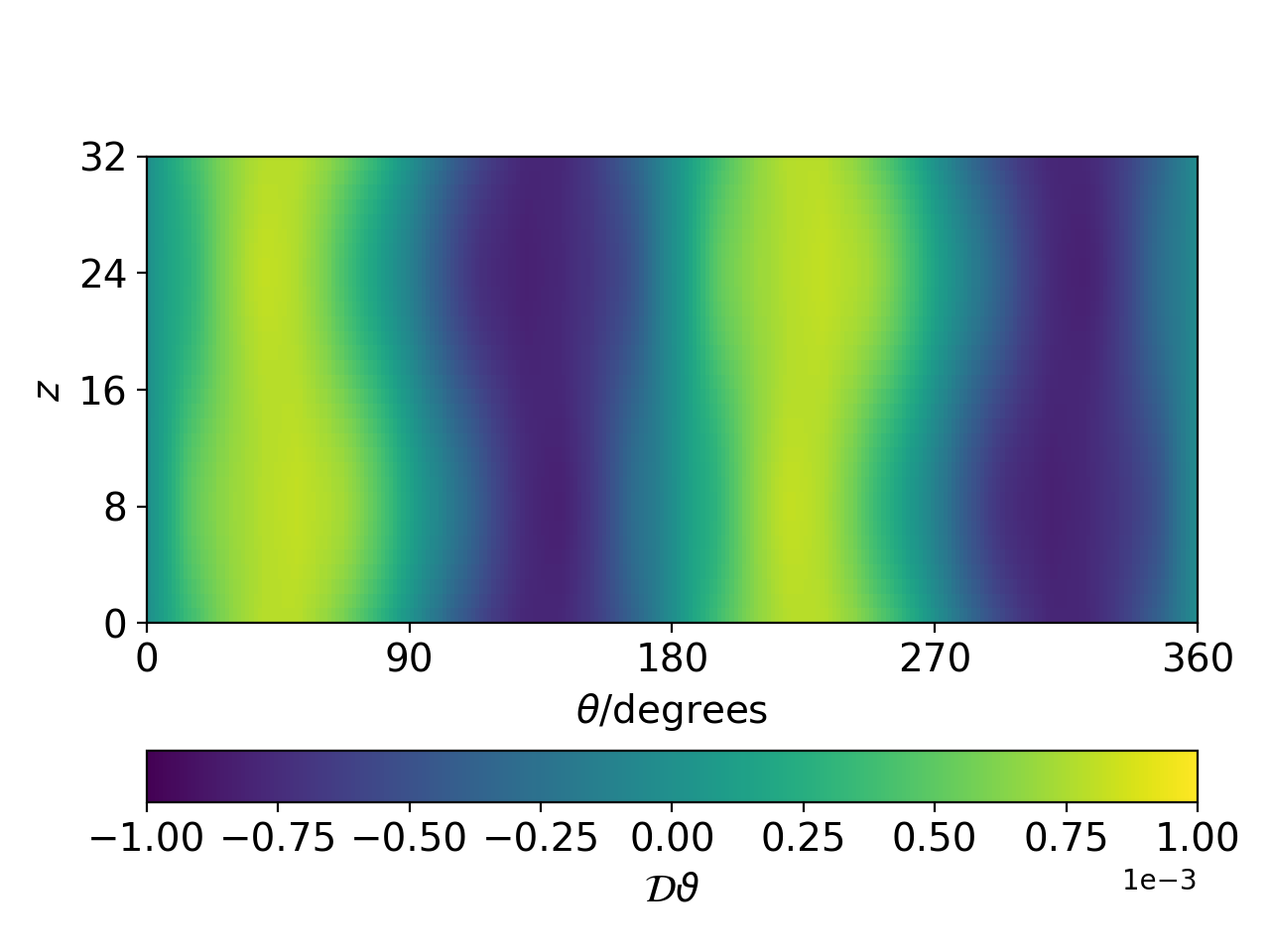}
    \caption{Massless radiation ${\cal{D}}\vartheta \cdot \hat{\bf r}$ from a $\lambda=1$ string with initial amplitude $A_0=1$  at late time $t=167.5$, measured on a cylinder at $R=64$, where $\theta$ is the azimuthal angle. The dominant quadrupole mode $\{mn\} = \{2\,0\}$ can be clearly identified.}
    \label{SmallAmplitudet670Lambda1Massless}
\end{figure}

\begin{figure}
    \centering
    \includegraphics[width=0.48\textwidth]{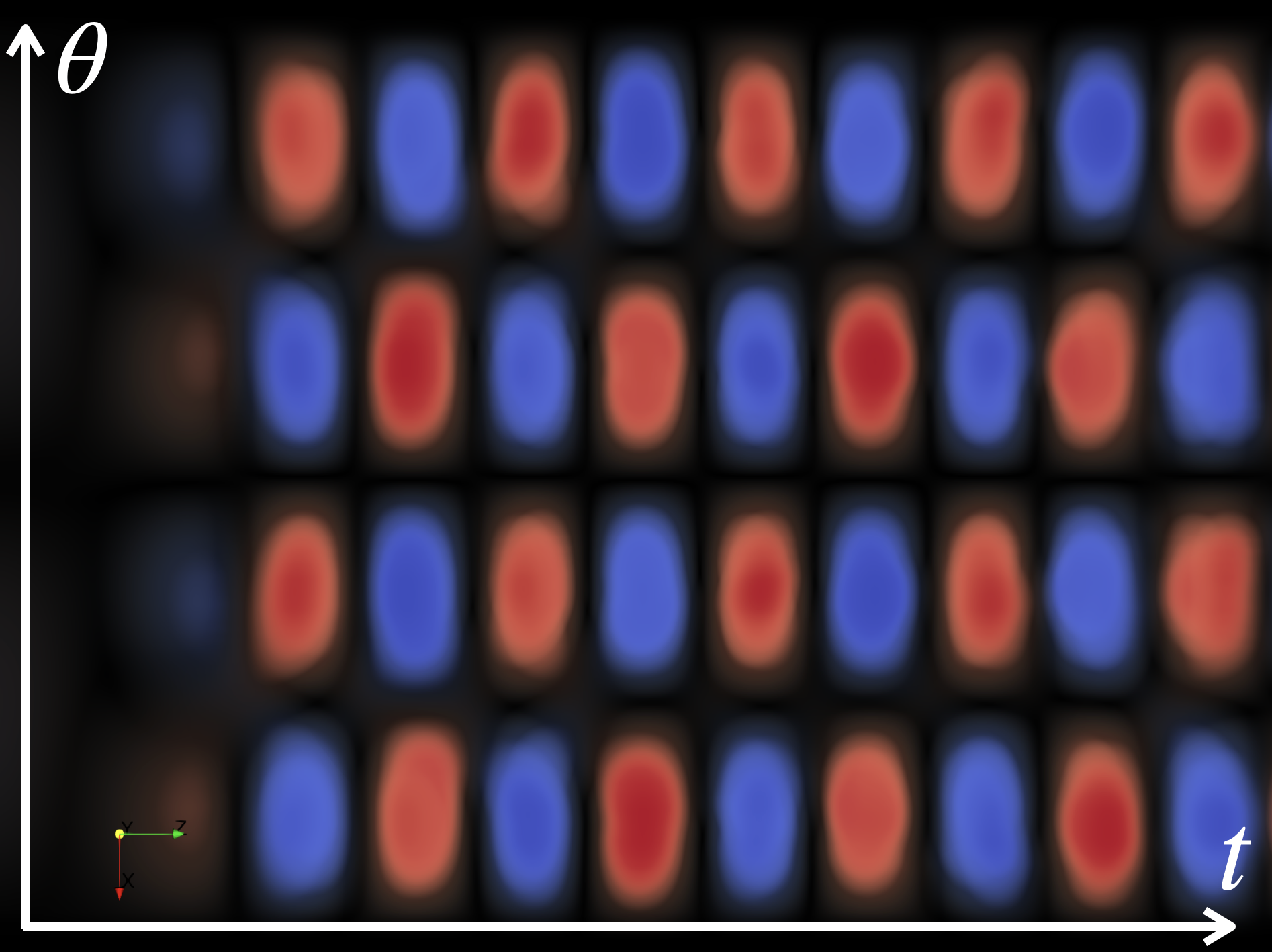}
    \caption{Volume rendering in spacetime $(t, \theta, z)$ of the massless radiation ${\cal{D}}\vartheta \cdot \hat {\bf r}$ from a $\lambda = 10$ string with initial amplitude $A_0=1$ over time, measured on a cylinder at $R = 64$. The time axis runs left to right, the azimuthal angle $\theta$ from bottom to top and the $z$-axis out of the page. The dominant quadrupole mode $\{pmn\} = \{2\,2\,0\}$ can be clearly identified.}
    \label{2DRadiationField}
\end{figure}

\subsubsection{Small Amplitude Oscillations}


Here we present results of string simulations with small initial amplitude $A_0=1$ ($\varepsilon = 0.20$) for $\lambda=1$ and $\lambda=10$. We first present qualitative results from visualisation of the radiation extracted on the diagnostic cylinder. Figure \ref{SmallAmplitudet670Lambda1Massless} shows the massless radiation field extracted for one timestep at late time at $R=64$, revealing an $m=2$ angular dependence and, to a first approximation, no $z$-dependence (i.e. $n=0$). In Figure \ref{2DRadiationField}, the radiation field on the cylinder is plotted as a function of both space and time (2+1D), showing the consistent  periodic behaviour of the propagating field. From the time-dependence, we can infer this mode to be a second harmonic of the fundamental period ($p=2$), so that the observed quadrupole corresponds to the $\{pmn\} = \{2\,2\,0\}$ eigenmode from the asymptotic general solution (\ref{generalsolution}).

\begin{figure}
    \centering
    \includegraphics[width=0.47\textwidth, trim=0 70 0 0, clip]{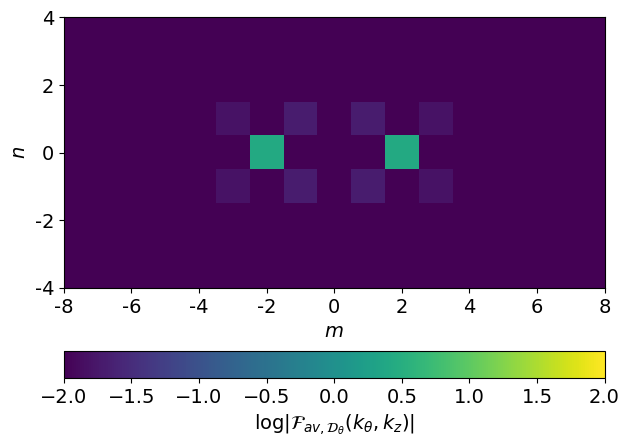}
    \includegraphics[width=0.47\textwidth, trim=0 70 0 10, clip]{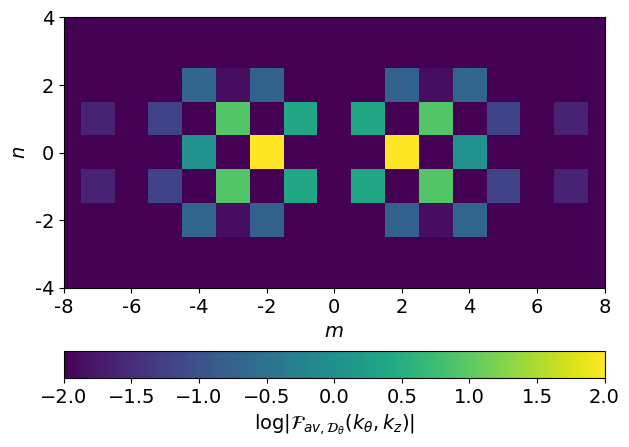}
     \includegraphics[width=0.47\textwidth, trim=0 0 0 10, clip]{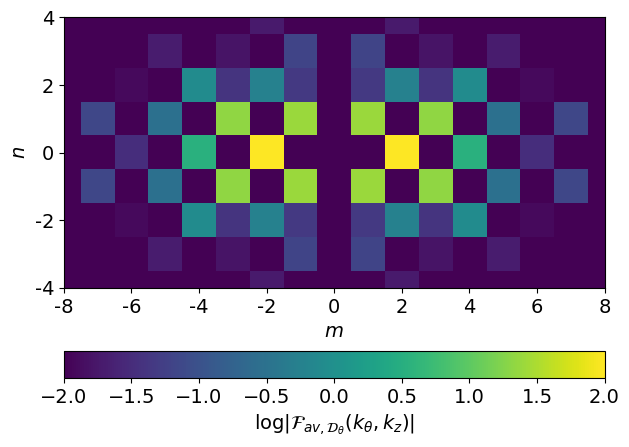}
   \caption{2D Fourier eigenmodes of the massless radiation ${\cal{D}}\vartheta \cdot \hat{\bf r}$ from a $\lambda=1$ string at late time $t=140.75$, measured on a cylinder at $R=64$ and time averaged over approximate half-period $\Delta t/2 = 66/4$. The horizontal axis is the angular eigenvalue $m$, while the vertical is the $z$-dependent wavenumber $n$. The top figure is for an initial amplitude $A_0=1$, the middle is for intermediate $A_0=4$ and the bottom is large $A_0=8$.   In all cases, the quadrupole signal $\{pmn\} = \{2\,2\,0\}$ is dominant, but higher harmonics contribute at larger amplitudes, provided they satisfy the checkerboard selection rule: $m+n$ even. }
    \label{checkerboard}
\end{figure}
\begin{figure}
    \centering
    \includegraphics[width=0.5\textwidth]{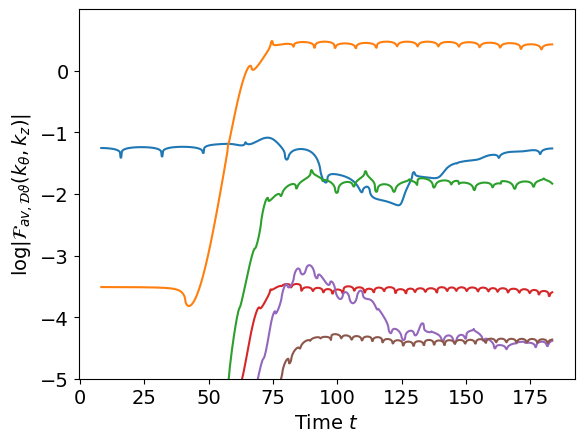}
    \includegraphics[width=0.5\textwidth]{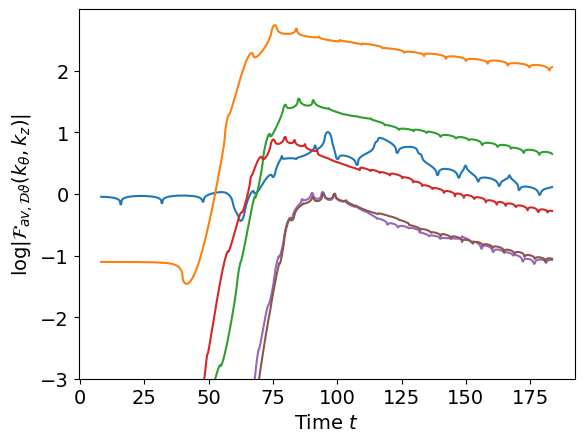}    
    \includegraphics[width=0.5\textwidth]{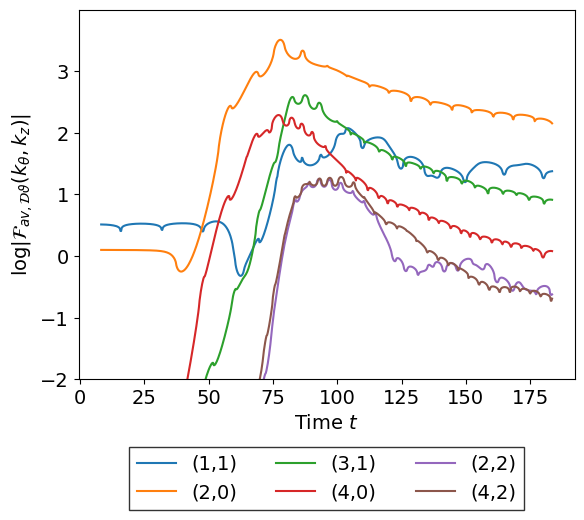}    

    \caption{Dominant 2D Fourier modes of the massless radiation ${\cal{D}}\vartheta \cdot \hat{\bf r}$ from a $\lambda=1$ string measured on a cylinder at $R=64$ and time averaged over approximate half-period $\Delta t/2 = 66/4$. The top figure is for an initial amplitude $A_0=1$, the middle is for intermediate $A_0=4$ and the bottom is large $A_0=8$.}
    \label{masslessspatialamp1regridding0.25cumulative}
\end{figure}

The extracted radiation field can be quantitatively analysed by decomposing into its constituent 2D Fourier modes $\mathcal{F}_{{\cal D}\vartheta}(k_\theta,k_z)$ using a 2D Fast Fourier transform (FFT). We can average these eigenmode signals over time to obtain a measure of their overall magnitude using
\begin{equation}\label{modes}
\mathcal{F}_{\mathrm{av}, {\cal D}\vartheta}(k_\theta,k_z) = \sum^{t = \Delta t/4}_{t = -\Delta t/4}{2\,\mathcal{F}_{{\cal D}\vartheta}(k_\theta,k_z)} / \Delta t\,,
\end{equation}
where $\Delta t$ is approximately one period of oscillation. Figure \ref{checkerboard} plots the pattern of $\{mn\}$ eigenmodes $\mathcal{F}_{\mathrm{av}, {\cal D}\vartheta}$ for $\lambda=1$ extracted at late time $t=140.75$, at which point the propagating massless signals have reached the cylinder. We obtain a `checkerboard' pattern that confirms the analytic selection rule discussed in Section \ref{dualradiation}, i.e. that only $m+n$ {\it even} eigenmodes can be generated. From Figure \ref{checkerboard}, we determine the six highest magnitude propagating modes, for which the time-average is plotted over time in the top panel of Figure \ref{masslessspatialamp1regridding0.25cumulative} for $\lambda=1$. A logarithmic scale is employed to highlight the  separation in amplitudes between the harmonics. We conclude that the quadrupole $\{2\,2\,0\}$ eigenmode offers the most significant radiation pathway, as predicted analytically by (\ref{sineradiation}). We see from Figure \ref{masslessspatialamp1regridding0.25cumulative} that the next strongest propagating mode is the third harmonic dipole $\{3\,1\,1\}$ which has an approximate relative amplitude of $0.09$, corresponding to a relative energy loss below 1\% that of the quadrupole. We also note that the later arrival of the $\{3\,1\,1\}$ mode is consistent with $\sim$5\% lower propagation velocity, as predicted by (\ref{dispersionrelation}).    

\begin{figure}[!]
    \centering
    \includegraphics[width=0.48\textwidth]{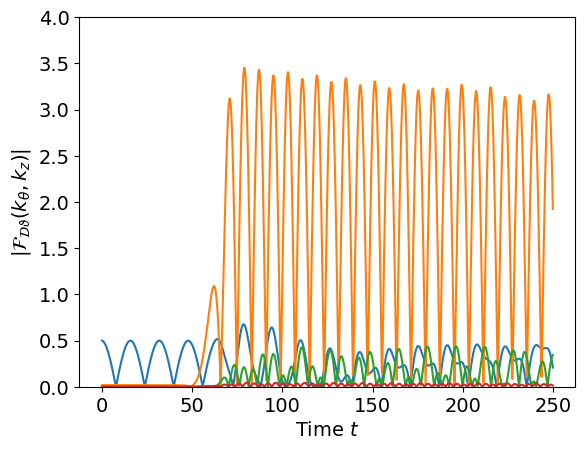}
    \includegraphics[width=0.48\textwidth]{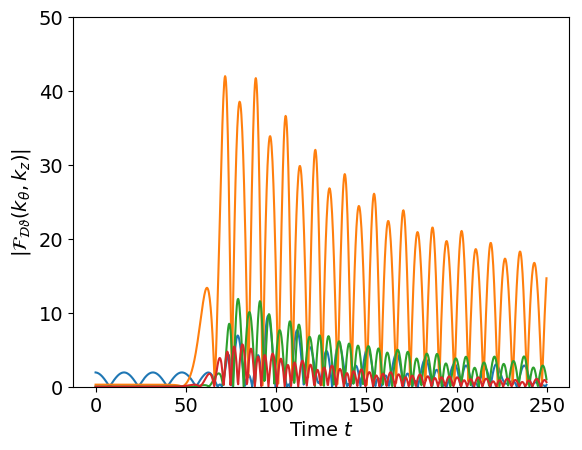}    
    \includegraphics[width=0.5\textwidth]{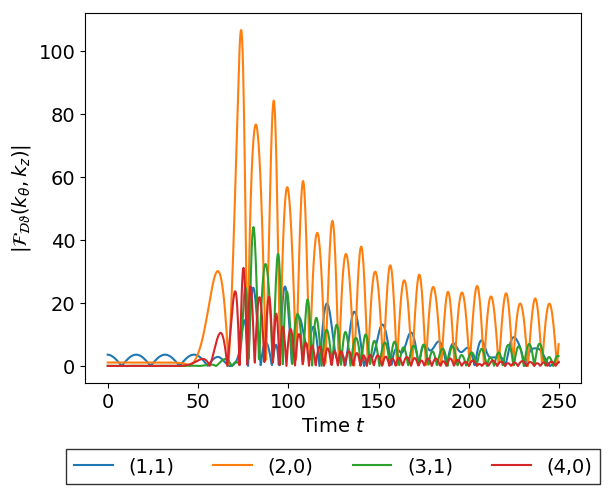}    
    \caption{Absolute value of the $\{mn\} = \{1\,1\}, \{2\,0\}, \{3\,1\},$ and $\{4\,0\}$ Fourier modes of the massless radiation ${\cal{D}}\vartheta \cdot \hat{\bf r}$ from a $\lambda=1$ string measured on a cylinder at $R=64$. The top figure is for an initial amplitude $A_0=1$, the middle is for intermediate $A_0=4$ and the bottom is large $A_0=8$. }\label{masslessspatialamp1regridding0.25}
\end{figure}

\begin{figure}[!]
    \centering
    \includegraphics[width=0.48\textwidth]{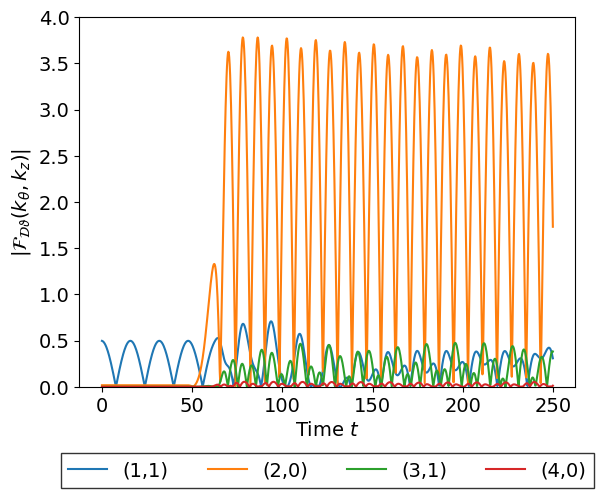} 
    \caption{Absolute value of the $\{mn\} = \{1\,1\}, \{2\,0\}, \{3\,1\},$ and $\{4\,0\}$ Fourier modes of the massless radiation ${\cal{D}}\vartheta \cdot \hat{\bf r}$ from a $\lambda=10$ string with initial amplitude $A_0=1$, measured on a cylinder at $R=64$.}\label{masslessspatialamp1regridding0.25lambda10}
\end{figure}

The absolute value of the four largest eigenmodes, $\{1\,1\,1\}$, $\{2\,2\,0\}$, $\{3\,3\,1\}$ and $\{4\,4\,0\}$, is shown in the top panel of Figure~\ref{masslessspatialamp1regridding0.25} for $\lambda=1$, where the time eigenvalue $p$ is inferred from the time period. (We note that the initial $\{1\,1\,1\}$ mode later acquires a small $\{2\,1\,1\}$ contribution.) The equivalent data for $\lambda=10$ is plotted in Figure~\ref{masslessspatialamp1regridding0.25lambda10}. The amplitudes and spectra of massless radiation for the two string widths with $\lambda=1$ and $\lambda=10$ are very similar.  However, some subtle differences are discernible, including a smaller initial radiation amplitude for $\lambda=1$ and a slightly faster amplitude decay rate.  Regarding the former, the initial massless radiation amplitudes are expected to be the same for all $\lambda$ (with which we will see our results agree for $\lambda \gtrsim 3$). However, for $\lambda=1$ with $A_0=1$ and $L=32$, finite size effects become important as the string core with $\phi < 1$ extends into the radiation zone (here, around $R\lesssim 4$), causing some suppression of the quadrupole amplitude (see Figure \ref{StringCrossSection}). The latter is a consequence of the $\lambda=1$ string being lighter, so there is a larger relative effect from radiation backreaction, as we will discuss in the next section.

We finally note that the dipole mode $\{1\,1\,1\}$ is present from the beginning of the simulation before radiation has had time to propagate to the cylinder, indicating that it is a long-range self-field of the oscillating string. As discussed in Section \ref{Massless}, this can be understood from the offset motion of the oscillating string fields from the centre of the diagnostic cylinder.  This apparent  $\{1\,1\,1\}$  wave  does not propagate, and so there should be no net flux over one period (if the amplitude remains constant). Using the spatial radiation diagnostic ${\cal D}\vartheta \cdot \hat{\bf r}$, the dipole self-field appears with an amplitude of $0.15$ relative to the propagating quadrupole mode. 



\begin{figure}
    \centering
    \includegraphics[width=0.5\textwidth]{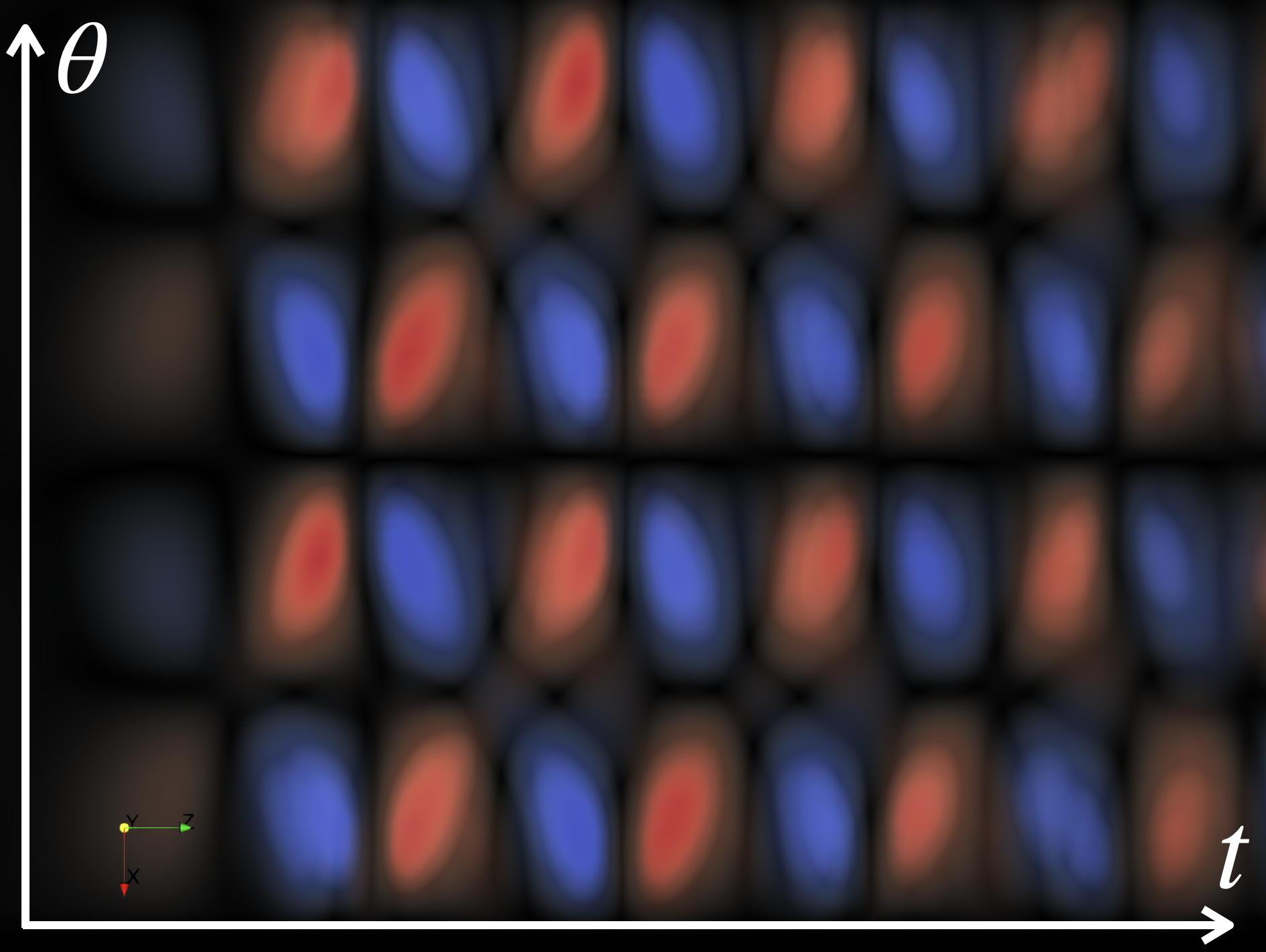}
    \caption{Volume rendering in spacetime $(t, \theta,z)$ of the massless radiation ${\cal{D}}\vartheta \cdot \hat{\bf r}$ from a $\lambda = 10$ string with initial amplitude $A_0=4$ over time, measured on a cylinder at $R = 64$. The time axis runs left to right and azimuthal angle $\theta$ from bottom to top and the $z$-axis out of the page. The dominant quadrupole mode $\{pmn\} = \{220\}$ can be clearly identified, but is distorted by higher modes.}
    \label{ParaviewMasslessScreenshotAmp4}
\end{figure}

\subsubsection{Large Amplitude Oscillations}

Here we present massless radiation results for larger initial amplitudes $A_0=4$ ($\varepsilon = 0.68$) and $A_0=8$ ($\varepsilon \approx 1$), with string widths given by $\lambda=1$ and $\lambda=10$. Figure \ref{ParaviewMasslessScreenshotAmp4} shows a visualisation of the radiation field measured on the diagnostic cylinder at $R=64$, measured over time, for the example $\lambda=10$ and $A_0=4$. Although the $\{2\,2\,0\}$ quadrupole mode remains dominant, the signal is modulated by higher harmonics. This is also illustrated in the lower panels in Figure \ref{checkerboard}, where many more modes are excited for $A_0\ge 4$ than for $A_0=1$ (upper panel). 

The centre panels of Figures \ref{masslessspatialamp1regridding0.25cumulative} and \ref{masslessspatialamp1regridding0.25} show the time-averaged and absolute magnitude of the largest propagating eigenmodes, $\{2\,2\,0\}$, $\{3\,3\,1\}$ and $\{4\,4\,0\}$ for $A_0=4$ and $\lambda=1$, as well as the self-field $\{1\,1\,1\}$ which is now mixed with the propagating dipole $\{2\,1\,1\}$. We see again that the $\{2\,2\,0\}$ quadrupole mode is dominant, contributing most of the outgoing radiation flux integrated across all modes.  Even in the highly nonlinear regime with $A_0=8$ ($\varepsilon \approx 1$) and $\lambda=1$ shown in the lower panel of Figure~\ref{masslessspatialamp1regridding0.25}, the next harmonic $\{3\,3\,1\}$ has a maximum relative amplitude $0.42$, i.e.\ initially contributing 18\% of the quadrupole energy flux, with $\{4\,4\,0\}$ around 8\% and $\{2\,1\,1\}$ 5\%.  We also note that the maximum quadrupole amplitude scales approximately with the relative oscillation amplitude squared $\varepsilon^2$, in agreement with expectations from (\ref{sineradiation}) that the energy flux scales as $\varepsilon^4$. Again, the amplitude of radiation from the lighter $\lambda =1$ string always decays more rapidly than the  $\lambda=10$ string because they initially have the same massless radiation output, a backreaction effect we shall discuss in the next section. We note that the amplitude decay of high harmonics ($n>2$) is considerably faster than the quadrupole, as illustrated in the lower panel of Figure~\ref{masslessspatialamp1regridding0.25cumulative}.


\subsection{String Radiation Backreaction}

\begin{figure*}
    \centering
     \includegraphics[width=0.75\textwidth]{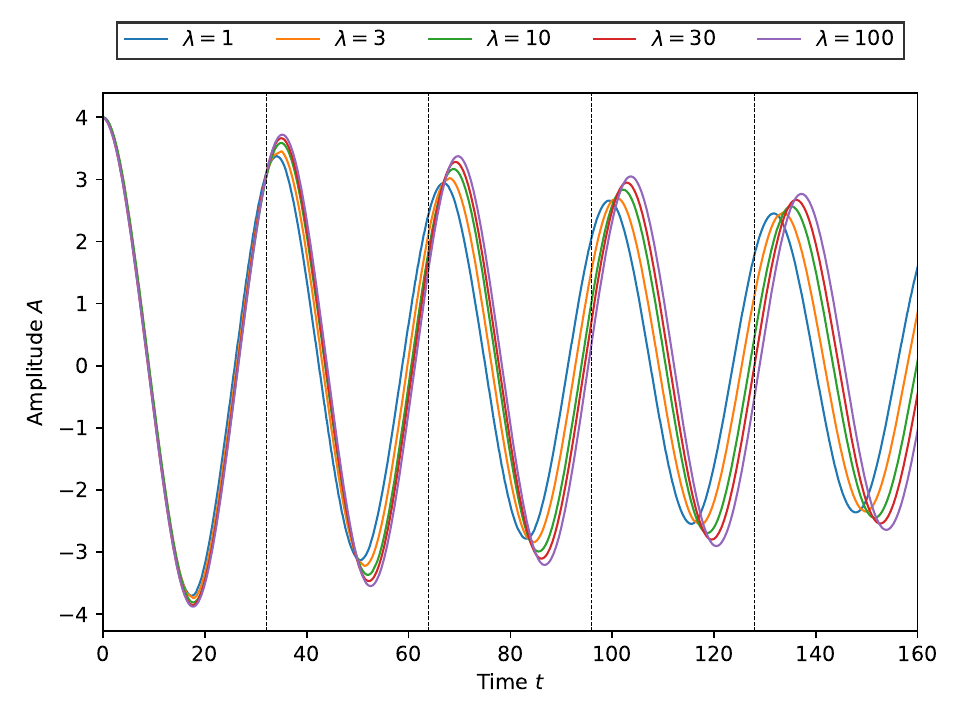}
    \caption{String amplitude over time for a selection of $\lambda$ parameter values in the range $1\le\lambda\le100$ with an initial amplitude $A_0=4$. The amplitude is measured at the point of maximum displacement $z = L/4$. The black dashed lines are plotted at intervals of $t=32$.}
    \label{amplitudedecay}
\end{figure*}

In this section, we analyse the detailed evolution of the oscillating string trajectories, observing the decay in amplitude due to radiation backreaction and comparing with analytic model predictions.  Focussing on regimes where the AMR evolution is robust and accurate, we analyse two specific sets of string simulations with amplitudes $A_0 = 1$ and $A_0= 3$ ($\varepsilon = 0.20, \, 0.54$), varying the string width parameter $\lambda$ across the wide range $1\le \lambda \le 100$.  We note that in the present AMR implementation, large amplitude $A_0\gtrsim 4$ ($\varepsilon \gtrsim 0.7$) oscillations at $\lambda \gtrsim 3$ (such as those illustrated in Figure \ref{amplitudedecay}) appear to be susceptible to small cumulative grid refinement effects at late times, which may have an effect on the evolution, as discussed in Section \ref{Discussion}.


We first plot string trajectories over time for a representative sample of $\lambda$ in Figure \ref{amplitudedecay}, which shows the decay of the string amplitude. The amplitude is taken to be the position of the string core at the $z$-coordinate of maximum string displacement, $z = N_3/4 \equiv L/4$, calculated using the winding algorithm described in Section \ref{stringcoreposition}. We see that as $\lambda$ increases, the rate of decay of the string generally decreases, indicating weaker radiation backreaction on strings with larger $\mu$. We also observe that the period of oscillation of the string $T>L$, as outlined in Section \ref{separableradiation}. Finally, we note that the period for the lightest string $\lambda=1$ approaches $T=32$ as the amplitude falls, where the higher mass strings remain at a higher periodicity. This is due to the larger radiation backreaction for lighter strings.

\begin{figure}
    \centering
    \includegraphics[width=0.5\textwidth]{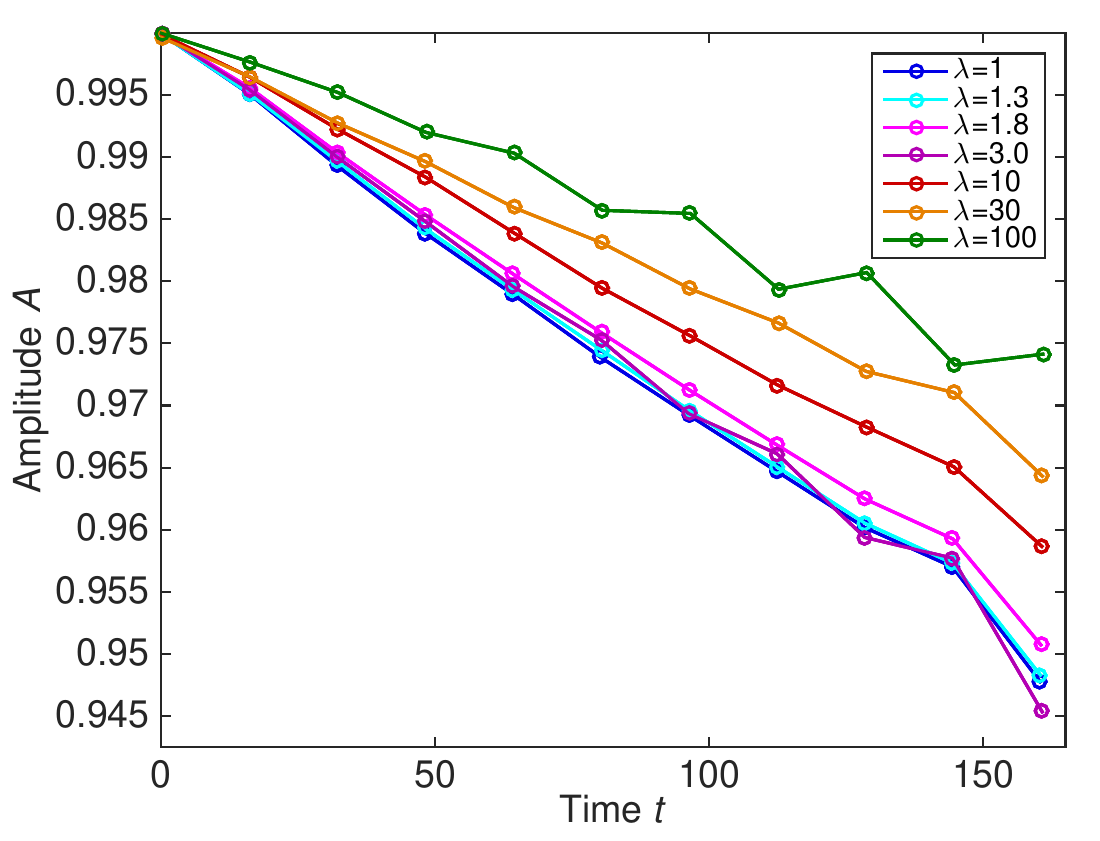}
    \includegraphics[width=0.5\textwidth]{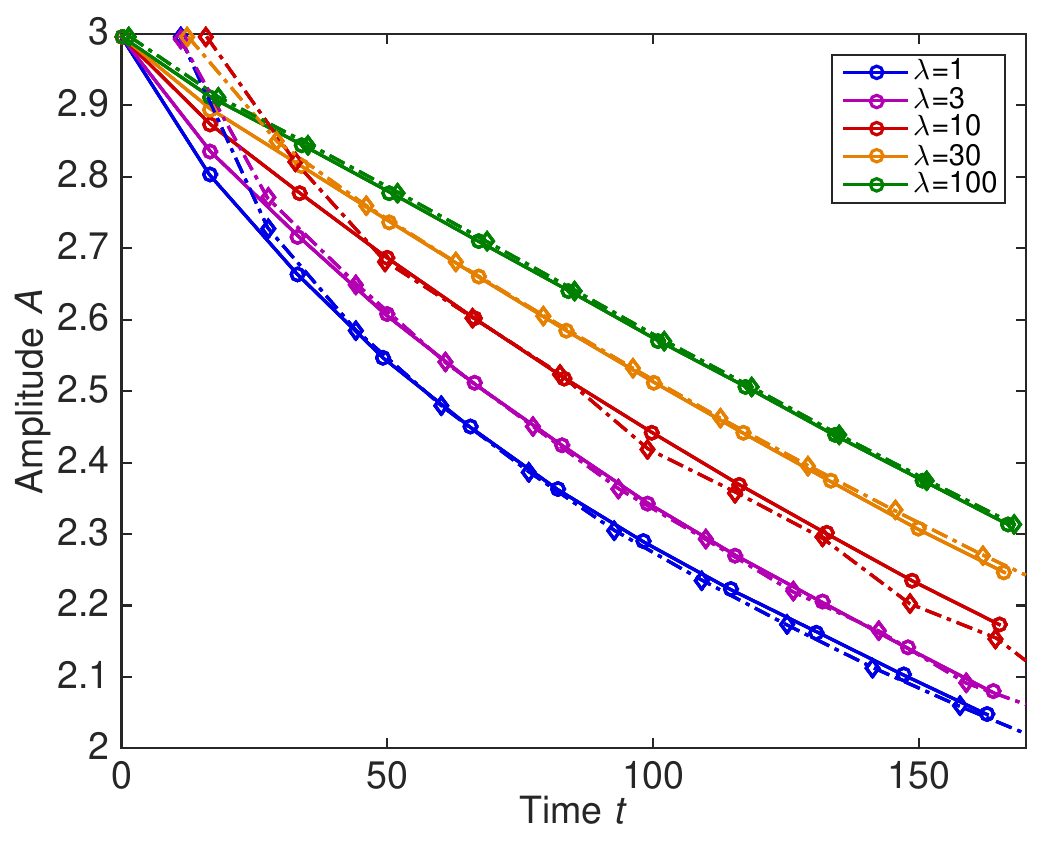}
    \caption{Maximum amplitude of decaying string oscillations for small amplitude $A_0=1$ (top) and intermediate amplitude $A_0=3$ (bottom) for different $\lambda$ at fixed length ($L=32$). In both cases, thin global strings (large $\lambda$) have a slower decay rate as the energy density $\mu$ is higher. At small $\lambda < 3$, radiative decay reduces because of finite width effects. For $A_0=3$, additional data is plotted (diamond and dashed lines) for strings with initial conditions after enhanced relaxation. Here, the initial decay rate is faster but the asymptotic radiative decay is the same (as shown with appropriate time translations). The simple backreaction model \eqref{inversesquare} predicts that the linear slope depends on the string energy density (effectively $\ln \lambda$), but is independent of the amplitude $A$.}
    \label{fig:decayA1}
\end{figure}


We can model the rate of decay by extracting the maximum and minimum amplitude of the string for each period of oscillation. Figure \ref{fig:decayA1} plots these values for the two data sets $A_0 = 1$ and $A_0= 3$.  The extrema of the oscillating string with small amplitude ($\varepsilon = 0.20$) shown in the top panel reveal nearly linear decay with a weak damping rate that decreases, as expected, with increasing $\lambda$ (i.e. as the effective mass per unit length of the string increases).   However, at small $\lambda\lesssim 3$, the radiative decay stalls and it becomes difficult to distinguish different $\lambda$.  In this regime, the oscillation amplitude $A_0=1$ is very close to the string width $\delta = 1/\sqrt\lambda \gtrsim 0.6$, where massive internal excitations within the string core can be expected to represent a non-negligible part of any string oscillation. These `breather' modes mean that the motion of the zero ($\phi=0$) at the string core is likely to be larger than the actual centre of mass oscillation, where the center of mass is determined by the motion of the dominant massless fields from which the radiation emanates. In our subsequent analysis, we make a small correction for this finite width effect. (We also note that the $\lambda=100$ string with $A_0=1$ appears to have drifted slightly from the centre from which the maximum amplitude is measured. This is due to the small difference in amplitude of the quadrupole radiation produced when the string is moving in the forward or backward direction relative to the propagation direction.)

The bottom panel of Figure \ref{fig:decayA1} shows maxima from string oscillations of intermediate amplitude ($\varepsilon=0.54$), showing trajectories with significant curvature, especially for the lighter strings (small $\lambda$) with more damping.  This figure also illustrates the effect of different initial conditions due to changing the timescale of preceding relaxation before releasing the string to undergo relativistic hyperbolic evolution (see Section \ref{globalstrings}).   The second set of data points (dotted lines) shows `over-relaxed' initial conditions where the gradient flow phase was started much earlier, thus removing longer-range correlations.  This hastens the initial amplitude decay but, asymptotically, the radiating string settles into a steady state which closely matches that from the other initial conditions, as can be shown by a simple time translation. The `under-relaxed' case (not shown here) exhibits opposite behaviour with a smaller initial decay, but again the same asymptotic limit.  These simulations were also performed using grid refinement levels at which there was no discernible improvement from increasing refinement further. 

\subsubsection{Inverse square amplitude model}

\begin{figure}[!]
    \centering
    \includegraphics[width=0.5\textwidth]{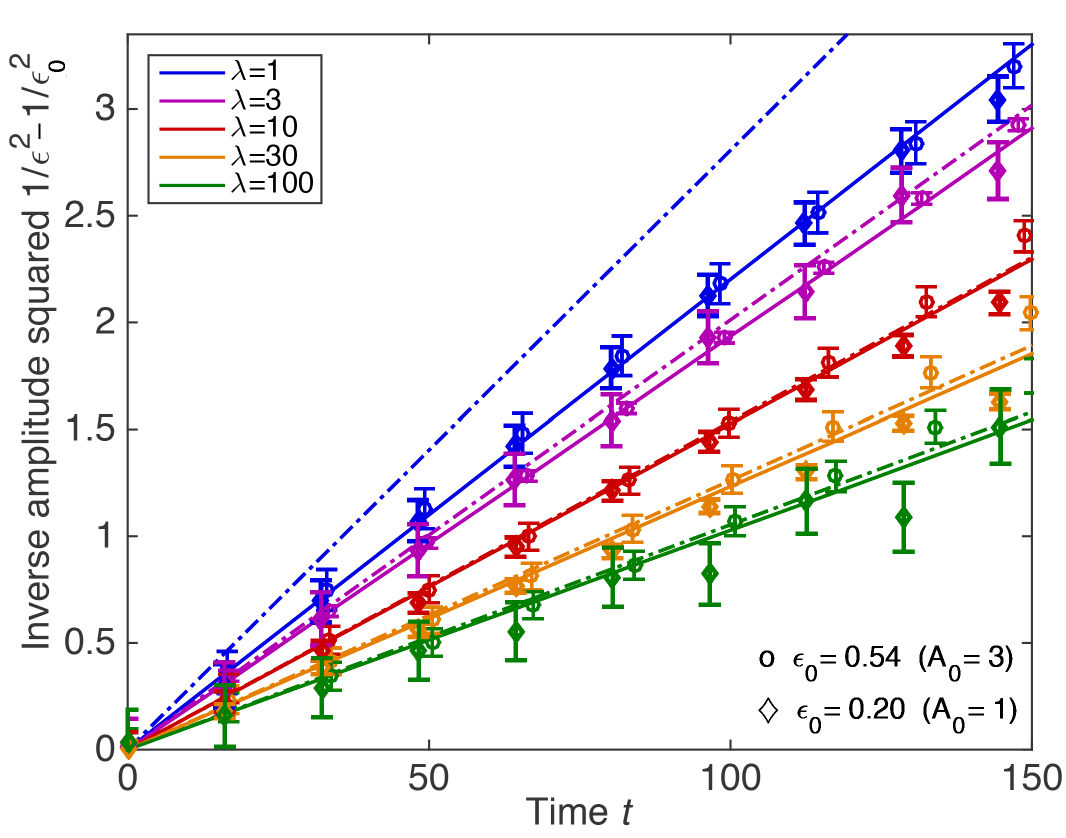}
    \caption{Fit of oscillating string data to the simple backreaction model (\ref{inversesquare}) for the inverse square of the relative amplitude $\varepsilon=2\pi A/T$ against time $t$.  After a small 8\% finite width correction, the $A_0=1$ ($\varepsilon=0.2$) data (diamonds) align closely with the larger amplitude data (circles), showing consistent linear behaviour for all $\lambda$.  The analytic prediction from the inverse square model (\ref{inversesquare}) for the best fit parameters is plotted (solid lines for $A_0=3$, dashed lines for $A_0=1$) for each $\lambda$, showing good agreement for all $\lambda >3$.}
    \label{fig:inv_square_fit}
\end{figure}

Analytic radiation calculations for a sinusoidal oscillatory string (\ref{sineapprox}) yield a specific prediction for the backreaction effect on the string trajectory (\ref{inversesquare}); the inverse square amplitude $1/A^2$ (or $1/\varepsilon^2$) is predicted to be linearly related to the time $t$. We find agreement with this for both data sets $A_0 = 1$ and $A_0= 3$, as shown in Figure~\ref{fig:inv_square_fit}, which plots the data for each $\lambda$ and both amplitudes superposed with zero intercept. Both datasets are proportional to $1/\varepsilon^2$ and independent of $A_0$ and $\lambda$ as predicted. The different $A_0$ have approximately matching slopes for the same $\lambda$ values, i.e. the string energy density $\bar \mu$ alone determines the damping rate (or, equivalently, $\bar \mu \sim \ln \lambda$).  We can understand this physically; given that the radiation power is independent of $\lambda$ as shown by (\ref{sineradiation}), it can be shown by straightforward manipulation that the greater oscillation energy (\ref{sineenergy}) of the heavier strings with large $\lambda$ causes the amplitude $\varepsilon$ to decay more slowly.  


Finite width effects for the fat lighter strings ($\lambda \lesssim 3$) at small amplitude $(\varepsilon= 0.20$) reduce the damping rate (slope) dependence on $\ln \lambda$, as discussed previously.  Assuming that internal modes (within the  string thickness $\delta$) imply that the true string oscillation amplitude is slightly smaller than measured for all $\lambda$, we apply a finite width correction (fwc) to all data by modifying the raw amplitude $A$ to a new value $A'$ as follows:
\begin{equation}
A' = A - \xi /\sqrt\lambda \,.
\end{equation}
A small correction $\xi = 0.08$ (i.e. only 8\% of the string width) aligns the respective slopes of the $A_0 = 1,\, 3$ data sets remarkably well.  We note that this small linear correction is not adequate to align the data with the backreaction model when $A_0 \sim \delta$ as for the $\lambda=1$ case with $\delta \sim 1$, where much larger deviations are evident. For this reason, we exclude the $\lambda=1$ string data from our asymptotic parameter estimates in the upcoming analysis. This is significant, as $\lambda=1$ is the case on which most previous numerical studies have been based. There is also some evidence for deviation from linear behaviour at late times for the $\lambda=100$ string at larger amplitude $A_0=3$. We eliminate these last few time points from the analysis, as this behaviour is unphysical and due to a systematic effect linked to the numerical evolution (see Section \ref{Discussion}).

\begin{figure}
    \centering
    \includegraphics[width=0.5\textwidth]{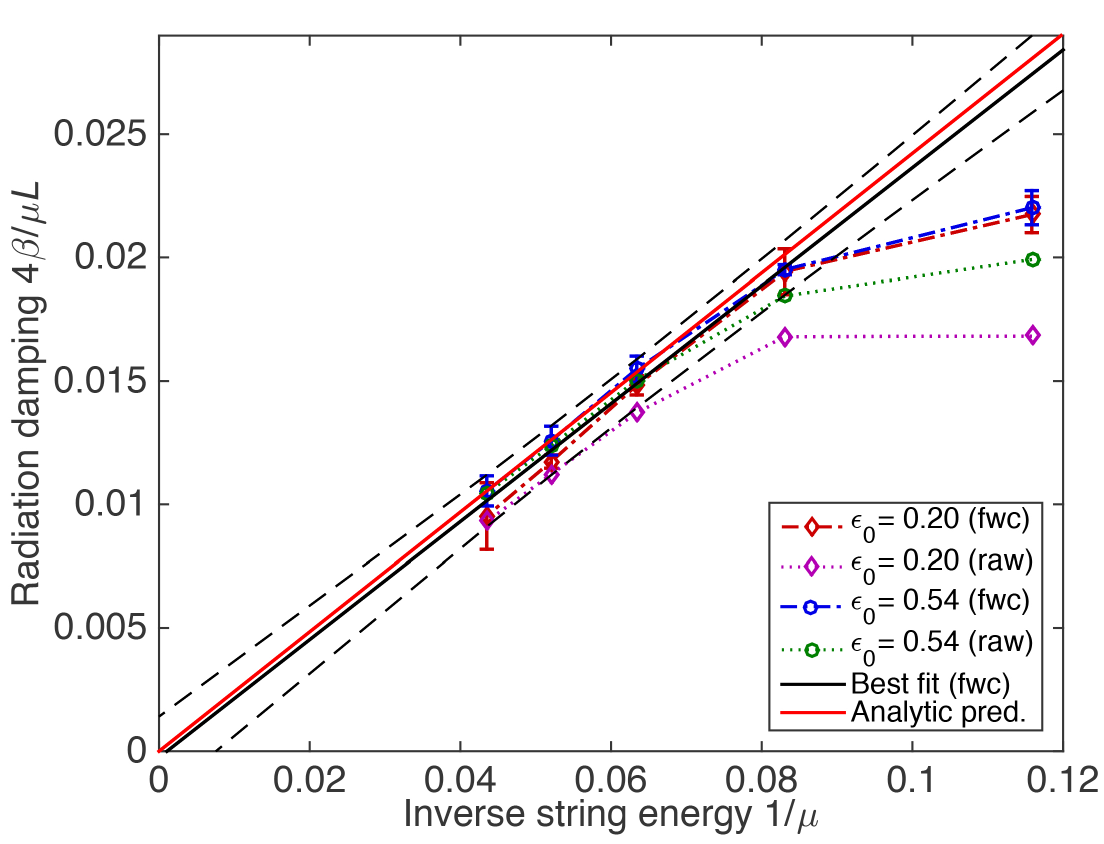}
    \caption{Measured radiative damping rates plotted as a function of inverse string density $\mu^{-1}$ (essentially the inverse $\ln \lambda$), also showing errors in the extrapolated slope.  Here we interpret the results with an effective string radius cutoff $R=3.5$, for which the damping rate vanishes as $\mu \rightarrow \infty$ ($\lambda \rightarrow \infty$).  We apply a finite size correction (fwc) and exclude the datapoint at $\lambda=1$, due to string width effects which limit radiative damping at small $\lambda$.}
    \label{fig:damping_vs_mu}
\end{figure}

In order to determine the accuracy of the inverse square backreaction model, we perform a least-squares best fit for each data set shown in Figure~\ref{fig:inv_square_fit} to determine the damping rate. The best fit lines are plotted, further illustrating the consistency between the two data sets at different amplitudes. We simultaneously estimate $\beta$ and $R$ for the backreaction model \eqref{inversesquare} from our data to determine the best fit. Figure~\ref{fig:damping_vs_mu} shows the string damping rate plotted against the string energy density $\bar\mu$.  We observe a key result: damping rates associated with $\lambda=3,10,30$ and $100$ align with the backreaction model when we take the cutoff scale $R\approx 3.75$, asymptotically projecting to zero damping as $\lambda \rightarrow \infty$, as expected for an infinitely heavy string. For this $R$, the analytic prediction (red line) shown in Figure~\ref{fig:damping_vs_mu} is in remarkable agreement, consistent with all damping rates, except those for $\lambda=1$ where finite size effects become important (for $L=32$). 

Despite this concordance with the inverse square model, there are fairly large uncertainties with a match possible within the parameter range
\begin{equation}
\beta =  7.6\pm 1.6\,, \quad \log R = 1.3 \pm 0.3 ~(R=3.75)\,.
\end{equation}
Without any finite width correction, the two data sets are less consistent, as reflected in larger uncertainties with $\beta = 9.5\pm 3.5$ and $\log R = 1.6\pm 0.6$ $(R=5.2)$.  These values for the radial string cutoff $R\approx 4$ may seem lower than those anticipated for a periodicity scale given by $L=32$.  However, the maximum radius of curvature for a large amplitude perturbation is $R\lesssim L/4 = 8$, so half this scale for the effective radius is not unreasonable.   Observing quadrupole radiation emanating from an oscillating global string heuristically indicates a delocalised process with radiation maxima appearing on a comparable scale to the perturbations on the string (see Figure~\ref{ParaviewMasslessScreenshot}).

We conclude that the analytic inverse square model (\ref{inversesquare}) offers an excellent description of an oscillating and radiating global string, predicting both the correct power law and magnitude of the radiation damping.  Our results are consistent with the analytic damping rate $\beta = \pi^3/4 \approx 7.75$ and indicate an effective radial cutoff for the string $R \approx  L/8$ ($\varrho \equiv R/L \approx 0.12 $), about half the string radius of curvature $L/4$.  We also observe that finite width effects suppress radiative power in massless modes at small amplitudes comparable to the string width $A \sim \delta$. Hence, by accurately fitting to the inverse square model, we have demonstrated that dual radiation predictions from the Kalb-Ramond action (\ref{duality}) are accurate as we asymptotically approach the small width regime.

\begin{figure}
    \centering
    \includegraphics[width=0.5\textwidth]{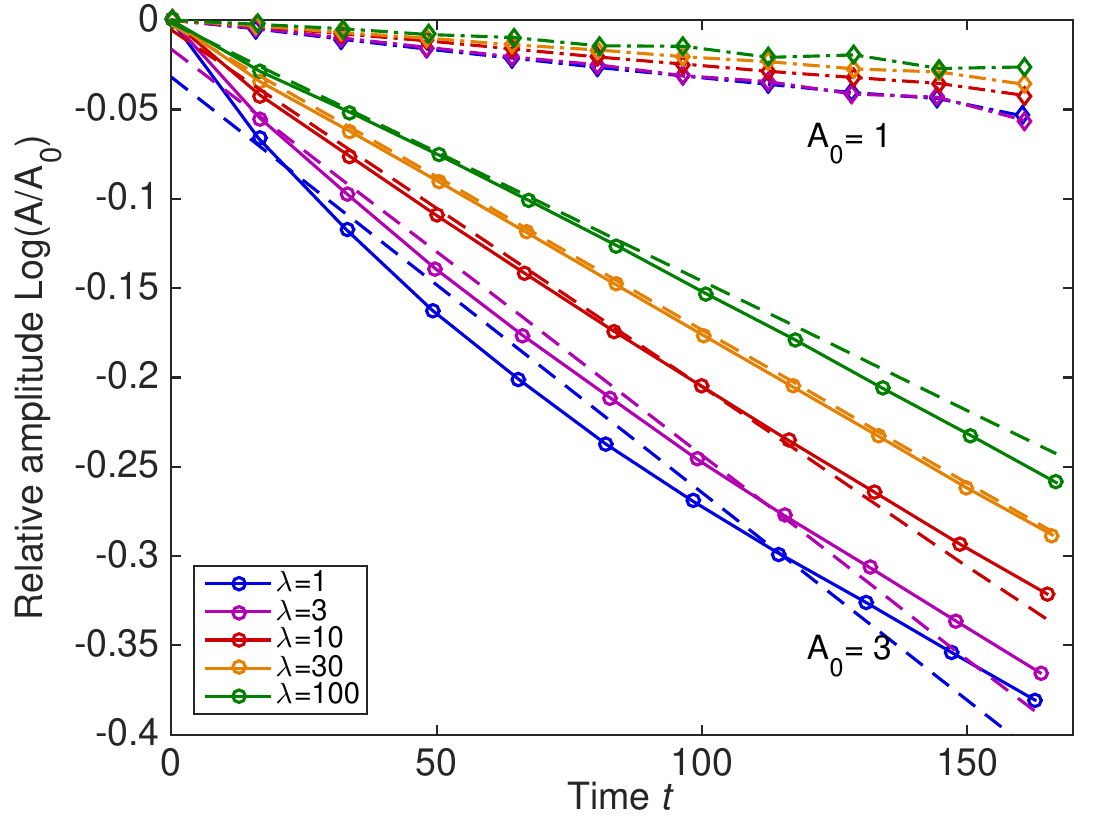}
    \caption{Relative amplitude $\ln(A/A_0) $ for oscillating strings as a function of time $t$.  The simple exponential decay model (\ref{exponential}) does not match the observed behaviour with the decay rate strongly dependent on the initial amplitude $A_0$ (contrast the inverse square model Figure~\ref{fig:inv_square_fit}).  The best fit lines (dashed) show clear deviations from exponential behaviour when there is larger damping at $A_0=3$.}
    \label{fig:exp_all}
\end{figure}

\subsubsection{Exponential Damping}

For comparison, we also endeavour to fit the oscillating string data for different $\lambda$ to the simple exponential damping model (\ref{exponential}). The analysis is plotted in Figure~\ref{fig:exp_all} using the logarithm of the relative amplitudes for different $\ln(A/A_0)$ as a function of time $t$.   Although the leading-order behaviour is linear at small amplitude $A_0=1$, as it is also for the inverse square model (\ref{inversesquare}), there are clearly significant deviations from exponential behaviour at larger amplitude $A_0=3$ (with the best fit exponentials deviating from the measured amplitudes).   More significantly, there is clearly a large decay rate dependence on the initial amplitude which is inconsistent with the simple model (\ref{exponential}).  Including an amplitude dependence $\varepsilon_0^2$ in the exponent improves the consistency of damping rates between the two amplitudes and so indicates that it may be applicable to situations with unequal left- and right-moving modes (see earlier discussion).   For our sinusoidal solution here, with equal left- and right-moving modes, it is clear that the inverse square model (\ref{inversesquare}) provides a better description of the observed damping behaviour.

\section{Conclusion \& Future Directions}\label{Conclusion}

We have presented results from the first fully adaptive mesh simulations of global cosmic strings, using the code GRChombo. We investigate single sinusoidally displaced string configurations with a wide range of string widths, defined by the parameter $\lambda$. The key purpose has been to obtain a robust asymptotic probe of the radiation emission from global or axion strings approaching cosmological scales, improving the limited dynamic range of previous numerical simulations for these configurations and comparing directly against dual radiation predictions using the Kalb-Ramond action (i.e.\ in the thin-string limit) \cite{Sakellariadou:1991sd,Battye1993}.  

We have studied massless (Goldstone boson or axion) radiation signals, using quantitative diagnostic tools to determine the eigenmode decomposition. As analytically predicted, the primary radiation channel for the sinusoidal string configuration is found to be the massless quadrupole eigenmode $\{2 \, 2\, 0\}$, completely dominating energy losses in all other modes ($n>2$) until we approach highly nonlinear configurations with relative amplitudes approaching unity, $\varepsilon \approx 1$.  Even in this nonlinear regime with a broader spectrum of eigenmodes present, the quadrupole remains the largest contribution, with backreaction rapidly suppressing the relative contribution from higher harmonics.  The massless radiation rate at a given small amplitude is independent of $\lambda$ for $L \gg \delta$, though finite size effects appear to cause some suppression around $\lambda \sim 1$ for our configurations ($L=32$). 

We have also compared oscillating string trajectories with the inverse square amplitude model \cite{Battye1993}, a backreaction model which accounts for radiation energy loss. Comparing with the analytically predicted amplitude decay rate, we show excellent correspondence for $\lambda > 3$.  Critically, the radiation damping rate depends inversely on the string tension $\mu = 2\pi \ln (R/\delta)$ which for a global string is renormalised by the string width $\delta \propto 1/\sqrt \lambda$ at a given curvature scale $R$.  Mitigating against finite width effects, we have been able to confirm the backreaction scaling law dependence $\mu^{-1}$ over the wide range $3\lesssim \lambda \lesssim 100$.  We conclude that global string evolution tends towards the behaviour predicted in the Nambu-Goto (thin-string) limit with radiation damping,  providing further confidence that analytic dual radiation modelling provides the appropriate large-scale (or cosmological) limit for global strings. 

The present work has implications for the study of axion radiation from global axion strings in the early universe, scenarios in which the Peccei-Quinn $U(1)$ symmetry is broken after inflation, forming a network of axion strings which decays as the axion mass becomes relevant \cite{Sikivie1982}.  As we have discussed, two  approaches have been used to calculate the number and spectrum of axions radiated by the string network; first, analytic radiation modelling combined with the results of Nambu string simulations and, secondly, direct numerical simulations of the underlying string field theory in an expanding universe.  Our present work on individual strings with a field theory study reaches higher numerical resolution than previous studies and offers some insight into the lack of agreement between these approaches.  Given that most network simulations to date use the comoving width (or `fat string') algorithm, they have an effective $\lambda\lesssim 1$ when compared to the configurations investigated here.   This is a regime where we have been able to identify a breakdown in correspondence with predictions from the thin-string limit and is also where light massive radiation channels begin to become competitive with massless radiation for nonlinear amplitudes.   

Our next step forward involves high resolution simulations of global string networks in an expanding background which are currently underway. By exploring different string widths with a range of $\lambda$ values and using our radiation diagnostics, we will endeavour to determine whether convergence towards the thin-string limit occurs and whether cosmological extrapolations are feasible numerically. These are important considerations which should reduce uncertainty in the present string predictions for the dark matter axion mass. This is also potentially relevant for predictions of gravitational waves from cosmic strings where there is even greater uncertainty, and where there is a close correspondence between string calculations for gravitational radiation and those for axion or antisymmetric tensor fields which have been tested in the present work. 
 
\section*{Acknowledgements}

We are grateful for useful conversations with Eugene Lim, Katy Clough, Thomas Helfer, Josu Aurrekoetxea, Miren Radia and Ulrich Sperhake. We would also like to thank and acknowledge the GRChombo team (http://www.grchombo.org/).

We are especially grateful to Kacper Kornet for invaluable technical computing support, continually offering patient advice and improving code parallelism, together with Alejandro Duran (Intel).  We would also like to acknowledge the support of the Intel Visualization team, led by Jim Jeffers, notably the collaboration on in-situ visualization with Carson Brownlee. 

We are further grateful for the organisers of the workshop `Cosmic Topological Defects: Dynamics and Multimessenger Signatures,' at the Lorentz Center, Leiden, for facilitating several helpful and enlightening discussions. AD would like to acknowledge networking support by the GWverse COST Action CA16104, `Black holes, gravitational waves and fundamental physics.'

This work was undertaken primarily on the COSMOS supercomputer at DAMTP, Cambridge, funded by BEIS National E-infrastructure capital grants ST/J005673/1 and STFC grants ST/H008586/1, ST/K00333X/1, as well as the Cambridge CSD3 part of the STFC DiRAC HPC Facility (www.dirac.ac.uk). The DiRAC component of CSD3 was funded by BEIS capital funding via STFC capital grants ST/P002307/1 and ST/R002452/1 and STFC operations grant ST/R00689X/1.

AD is supported by an EPSRC iCASE Studentship  in partnership with Intel (EP/N509620/1, Voucher 16000206).  EPS acknowledges funding from STFC Consolidated Grant ST/P000673/1.

\bibliography{Paper1new}

\end{document}